\documentclass[12pt]{iopart}

\usepackage{graphicx}
\begin{document}

\title[Casimir effects with 2D layers ]{Casimir effects in systems containing 2D layers, like graphene and 2D electron gases}

\author{Bo E. Sernelius}

\address{Division of Theory and Modeling, Department of Physics, Chemistry and Biology, Link{\"o}ping University, SE-581 83 Link{\"o}ping, Sweden}
\ead{bos@ifm.liu.se}

\begin{abstract}
We present a variety of methods to derive the Casimir interaction in planar systems containing two-dimensional layers. Examples where this can be of use is graphene, graphene-like layers and two-dimensional electron gases. We present results for two free standing layers and for one layer above a substrate. The results can easily be extended to systems with a larger number of layers.

\end{abstract}

\submitto{\JPCM special issue on Casimir Forces}
\maketitle
\section{Introduction}

The dispersion interaction is a correlation effect. The interaction is a result of the correlated motion of the charged carriers in different parts of the system. The dispersion energy between two objects is the inter-object correlation energy. In the field of condensed matter one often uses the Coulomb gauge. In Coulomb gauge the scalar potential is instantaneous and the vector potential is purely transverse. The scalar potential represents Coulomb interactions between the carriers while the vector potential represents photon interactions. If the two objects are not too far away from each other one may neglect retardation effects (effects from the finite speed of light in vacuum). Then only the Coulomb interactions contribute and the properties of the materials enter in terms of the density-density correlation functions. If retardation has to be taken into account the photon interactions have to be included and, in this additional contribution, the properties of the materials enter in terms of the current-current correlation functions. The interaction can be found using many-body theory\cite{Mah,Fet} with diagrammatic perturbation theory based on Feynman diagrams.

Alternatively the interaction can be viewed as caused by fluctuations, fluctuations in the charge and current densities or fluctuations in the electromagnetic fields. Then one way to find the interaction is in terms of the electromagnetic normal modes of the system.\,\cite{Ser1} These normal modes are massless bosons and at zero temperature the interaction energy is the sum of the zero-point energy of all these modes.

In a translational invariant and homogeneous system there are three types of normal mode; two with transverse fields and one with longitudinal fields. In the present system there are only two types of normal mode; one with $p$-polarized fields, i.e., with electric vector in the plane formed by the in-plane momentum $\bf k$ and the normal to the layers, experienced as longitudinal in the two-dimensional (2D) sheets; one with $s$-polarized fields, i.e., with electric vector perpendicular to this plane, experienced as transverse in the 2D sheets. The first type of mode is called TM (transverse magnetic) and second TE (transverse electric). The dispersion curves for each mode type can have one or several branches.

For planar structures the quantum number that characterizes the normal modes is {\bf k}, the 2D wave vector in the plane of the interfaces or planar structures and the interaction energy per unit area for each mode type can be written as

\begin{equation}
E = \frac{1}{A}\sum\limits_{\bf{k}} {\sum\limits_i {\frac{1}{2}\hbar \omega _{\bf{k}}^i} }  = \frac{1}{2}\hbar \int {\frac{{{d^2}k}}{{{{\left( {2\pi } \right)}^2}}}\sum\limits_i {\omega _{\bf{k}}^i} },
\label{equ1}
\end{equation}
where $A$ is the area of a planar interface and the summation over $i$ is a summation over the branches. To get the expression on the right-hand side we have let $A$ be large so that the summation over the discrete $\bf k$ variable can be replaced by an integral over a continuous $\bf k$ variable.

After using an extension of the so-called argument principle and deforming the integration path in the complex frequency plane one arrives at the expression\,\cite{Ser1}
\begin{equation}
E = \hbar \int {\frac{{{d^2}k}}{{{{\left( {2\pi } \right)}^2}}}} \int\limits_0^\infty  {\frac{{d\omega }}{{2\pi }}} \ln \left[ {{f_{\bf{k}}}\left( {i\omega } \right)} \right],
\label{equ2}
\end{equation}
where ${f_{\bf{k}}}\left( {\omega _{\bf{k}}^i} \right) = 0$ is the condition for electromagnetic normal modes. Equation\,(\ref{equ2}) is valid for zero temperature and the interaction energy is the internal energy. 

At finite temperature also the thermal population of the modes affects the interaction. The interaction energy is now Helmholtz' free energy and can be written as
\begin{equation}
\begin{array}{l}
E = \frac{1}{\beta }\int {\frac{{{d^2}k}}{{{{\left( {2\pi } \right)}^2}}}\sum\limits_{n = 0}^\infty  {'\ln \left[ {{f_k}\left( {i{\xi _n}} \right)} \right]} } ,\\
{\xi _n} = \frac{{2\pi n}}{{\hbar \beta }};\;n = 0,\,1,\,2,\, \ldots ,
\end{array}
\label{equ3}
\end{equation}
where $\beta  = 1/{k_B}T$. The integral over frequency has been replaced by a summation over discrete frequencies, the so-called Matsubara frequencies.\cite{Mah,Fet}  The prime on the summation sign indicates that the $n = 0$ term should be divided by two. 

The normal modes can be found in different ways. One way is to solve Maxwell's equations (in absence of any external sources) in all regions of the system and use the proper boundary conditions at the interfaces.
Another way is to find self-sustained solutions using induced charge- and current-densities in the objects.

We here derive the dispersion interaction in a variety of ways. Since the non-retarded treatment is much easier to perform we present both the non-retarded and fully retarded derivations in separate sections. In the derivations involving normal modes we derive the mode condition function, ${f_{\bf{k}}}\left( \omega  \right)$. The interaction energy is then found by inserting this function into (\ref{equ2}) or (\ref{equ3}). In the derivations involving diagrammatic perturbation theory the interaction energy is obtained directly.

We do not present any numerical results, only the final expressions and different ways to arrive at these. However, we give the analytical expressions for the dielectric function of graphene and of the 2D electron gas so it is straight forward for the readers to perform numerical derivations of their own. Numerical results for graphene by us and others can be found in the literature.\cite{Ser2,Ser3,Ser4,Ser5,Ser6,KliMosSer,Bordag2,Fial,Sara,Svet,Woods}

We begin in section \ref{NOT} with introducing our notation. The following two sections are devoted to background materials, electromagnetic in section \ref{EMBG} and 2D in section \ref{BG2D}. The main results when retardation effects are neglected are derived in section \ref{NonRet} and when retardation effects are taken into account in section \ref{Ret}. We end in section \ref{Sum} with a brief summary and conclusions.

\section{Notation\label{NOT}}
In this work we discuss planar structures. We let all surfaces, interfaces and 2D films be parallel to the $xy$-plane. We let 
${\bf{q}}$ be a three-dimensional (3D) wave vector and ${\bf{k}}$ its 2D in-plane component. The corresponding notation for the spatial coordinates are ${\bf{R}}$ and ${\bf{r}}$, respectively. Thus we have
\begin{equation}
\begin{array}{l}
{\bf{q}} = \left( {{q_x},{q_y},{q_z}} \right) = \left( {{k_x},{k_y},{q_z}} \right) = \left( {{\bf{k}};{q_z}} \right);\\
{\bf{R}} = \left( {{R_x},{R_y},{R_z}} \right) = \left( {{r_x},{r_y},z} \right) = \left( {{\bf{r}};z} \right);\\
{\bf{q}} \cdot {\bf{R}} = {\bf{k}} \cdot {\bf{r}} + {q_z}z.
\end{array}
\label{equ4}
\end{equation}
We will end up with a formalism only depending on ${\bf{k}}$ and ${\bf{r}}$ apart from the distances between the various interfaces. Thus the formalism is 2D.

We will make extensive use of the Fourier transform, both the 3D and 2D versions, and since they can be defined in different ways we list our definitions here. The 3D Fourier transform and inverse Fourier transform of a function $F$ with only spatial dependence are
\begin{equation}
\begin{array}{l}
F\left( {\bf{q}} \right) = \int {{d^3}R{e^{ - i{\bf{q}} \cdot {\bf{R}}}}F\left( {\bf{R}} \right)} ;\\
F\left( {\bf{R}} \right) = \int {\frac{{{d^3}q}}{{{{\left( {2\pi } \right)}^3}}}{e^{i{\bf{q}} \cdot {\bf{R}}}}F\left( {\bf{q}} \right)}.
\end{array}
\label{equ5}
\end{equation}
If the function also has a temporal dependence the corresponding relations are
\begin{equation}
\begin{array}{l}
F\left( {{\bf{q}},\omega } \right) = \int {{d^3}R\int\limits_{ - \infty }^\infty  {dt{e^{ - i\left( {{\bf{q}} \cdot {\bf{R}} - \omega t} \right)}}F\left( {{\bf{R}},t} \right)} } ;\\
F\left( {{\bf{R}},t} \right) = \int {\frac{{{d^3}q}}{{{{\left( {2\pi } \right)}^3}}}\int\limits_{ - \infty }^\infty  {\frac{{d\omega }}{{2\pi }}} {e^{i\left( {{\bf{q}} \cdot {\bf{R}} - \omega t} \right)}}F\left( {{\bf{q}},\omega } \right)}.
\end{array}
\label{equ6}
\end{equation}
In 2D the Fourier transform and inverse Fourier transform of a function are
\begin{equation}
\begin{array}{*{20}{l}}
{F\left( {\bf{k}} \right) = \int {{d^2}r{e^{ - i{\bf{k}} \cdot {\bf{r}}}}F\left( {\bf{r}} \right)} ;}\\
{F\left( {\bf{r}} \right) = \int {\frac{{{d^2}k}}{{{{\left( {2\pi } \right)}^2}}}{e^{i{\bf{k}} \cdot {\bf{r}}}}F\left( {\bf{k}} \right)} ,}
\end{array}
\label{equ7}
\end{equation}
if the function has only spatial dependence.
If the function also has a temporal dependence the relations are
\begin{equation}
\begin{array}{l}
F\left( {{\bf{k}},\omega } \right) = \int {{d^2}r\int\limits_{ - \infty }^\infty  {dt{e^{ - i\left( {{\bf{k}} \cdot {\bf{r}} - \omega t} \right)}}F\left( {{\bf{r}},t} \right)} } ;\\
F\left( {{\bf{r}},t} \right) = \int {\frac{{{d^2}k}}{{{{\left( {2\pi } \right)}^2}}}\int\limits_{ - \infty }^\infty  {\frac{{d\omega }}{{2\pi }}} {e^{i\left( {{\bf{k}} \cdot {\bf{r}} - \omega t} \right)}}F\left( {{\bf{k}},\omega } \right)}. 
\end{array}
\label{equ8}
\end{equation}
We will need yet another Fourier transform. We let the function $F$ be a function of the 3D spatial variable ${\bf{R}}$ but we only transform with respect to ${\bf{r}}$. Then we have for a function with only spatial dependence
\begin{equation}
\begin{array}{*{20}{l}}
{F\left( {{\bf{k}};z} \right) = \int {{d^2}r{e^{ - i{\bf{k}} \cdot {\bf{r}}}}F\left( {{\bf{r}};z} \right)} ;}\\
{F\left( {{\bf{r}};z} \right) = \int {\frac{{{d^2}k}}{{{{\left( {2\pi } \right)}^2}}}{e^{i{\bf{k}} \cdot {\bf{r}}}}F\left( {{\bf{k}};z} \right)} .}
\end{array}
\label{equ9}
\end{equation}
If the function also has a temporal dependence the relations are
\begin{equation}
\begin{array}{l}
F\left( {{\bf{k}};z,\omega } \right) = \int {{d^2}r\int\limits_{ - \infty }^\infty  {dt{e^{ - i\left( {{\bf{k}} \cdot {\bf{r}} - \omega t} \right)}}F\left( {{\bf{r}};z,t} \right)} } ;\\
F\left( {{\bf{r}};z,t} \right) = \int {\frac{{{d^2}k}}{{{{\left( {2\pi } \right)}^2}}}\int\limits_{ - \infty }^\infty  {\frac{{d\omega }}{{2\pi }}} {e^{i\left( {{\bf{k}} \cdot {\bf{r}} - \omega t} \right)}}F\left( {{\bf{k}};z,\omega } \right)}. 
\end{array}
\label{equ10}
\end{equation}
Note that we have refrained from introducing separate function names for the different Fourier transformed versions of a function.  By looking at the argument of the function we can tell if it is a 3D or 2D Fourier transform, a partial transform, or not a transform at all.
\section{Electromagnetic background material\label{EMBG}}
This section has two subsections. In the first, section \ref{ME}, we present the proper version of Maxwell's equations and also the boundary conditions at interfaces in the system. In the second, section \ref{POT}, we give the scalar and vector potentials in Coulomb and Lorentz gauges and show how the fields are obtained from these.
\subsection{Maxwell's equations\label{ME}}

There are different formulations of electromagnetism in the literature. The difference lies in how the conduction carriers are treated. In one formulation these carriers are lumped together with the external charges to form the group of free charges. Then only the bound charges contribute to the screening. We want to be able to treat geometries with metallic regions. Then this formulation is not suitable. In the formulation that we use the conduction carriers are treated on the same footing as the bound charges. Thus, both bound and conduction charges contribute to the dielectric function. In the two formalisms the {\bf E} and {\bf B} fields, the true fields, of course are the same. However, the auxiliary fields,  the {\bf D} and {\bf H} fields, are different. To indicate that we use this alternative formulation we put a tilde above the {\bf D} and {\bf H} fields and also above the dielectric functions, polarizabilities and conductivities.

With this formulation Maxwell's equations (MEs) have the form
\begin{equation}
\begin{array}{l}
\nabla  \cdot {\bf{\tilde D}} = 4\pi {\rho _{ext}}\\
\nabla  \cdot {\bf{B}} = 0\\
\nabla  \times {\bf{E}} + \frac{1}{c}\frac{{\partial {\bf{B}}}}{{\partial t}} = 0\\
\nabla  \times {\bf{\tilde H}} - \frac{1}{c}\frac{{\partial {\bf{\tilde D}}}}{{\partial t}} = \frac{{4\pi }}{c}{{\bf{J}}_{ext}},
\end{array}
\label{equ11}
\end{equation}
and the boundary conditions at an interface between two media $1$ and $2$ are
\begin{equation}
\begin{array}{l}
\left( {{{\bf{E}}_2} - {{\bf{E}}_1}} \right) \times {\bf{n}} = 0\\
\left( {{{{\bf{\tilde D}}}_2} - {{{\bf{\tilde D}}}_1}} \right) \cdot {\bf{n}} = 4\pi {\left( {{\rho _s}} \right)_{ext}}\\
\left( {{{\bf{B}}_2} - {{\bf{B}}_1}} \right) \cdot {\bf{n}} = 0\\
\left( {{{{\bf{\tilde H}}}_2} - {{{\bf{\tilde H}}}_1}} \right) \times {\bf{n}} =  - \frac{{4\pi }}{c}{{\bf{K}}_{ext}},
\end{array}
\label{equ12}
\end{equation}
where $\rho $ and ${\bf{J}}$ are volume charge and current densities, respectively, while ${\rho _s}$ and ${\bf{K}}$ are surface charge and current densities, respectively.
We note that the sources to the fields in MEs are the external charge and current densities and also in the boundary conditions discontinuities in the normal component of the ${\bf{\tilde D}}$ fields and tangential component of the ${\bf {\tilde H}}$ fields are caused by external surface charge densities and external surface current densities, respectively. The unit vector {\bf n} is the surface normal pointing into region $2$.

\subsection{Potentials\label{POT}}
Often it is easier to deal with potentials instead of the fields. The potentials are auxiliary functions, help functions. The potentials are not unique, they depend on which gauge we have chosen. Two very common gauges used in condensed matter theory are the Coulomb gauge (mentioned in the Introduction) and the Lorentz gauge. In Coulomb gauge the potentials are
\begin{equation}
\begin{array}{*{20}{l}}
{\Phi \left( {{\bf{R}},t} \right) = \int {\int {{d^3}R'dt'\frac{{\rho \left( {{{\bf{R}}^\prime },t'} \right)\delta \left( {t - t'} \right)}}{{\left| {{\bf{R}} - {{\bf{R}}^\prime }} \right|}}}  = \int {{d^3}R'\frac{{\rho \left( {{{\bf{R}}^\prime },t} \right)}}{{\left| {{\bf{R}} - {{\bf{R}}^\prime }} \right|}}} } ;}\\
{{\bf{A}}\left( {{\bf{R}},t} \right) = \frac{1}{c}\int {\int {{d^3}R'dt'\frac{{{{\bf{J}}_ \bot }\left( {{{\bf{R}}^\prime },t'} \right)\delta \left( {t - t' - \left| {{\bf{R}} - {{\bf{R}}^\prime }} \right|/c} \right)}}{{\left| {{\bf{R}} - {{\bf{R}}^\prime }} \right|}}} } .}
\end{array}
\label{equ13}
\end{equation}
Here the scalar potential, ${\Phi \left( {{\bf{R}},t} \right)}$, is instantaneous;  it describes the Coulomb interaction between the charged carriers; the Coulomb interaction is purely longitudinal. The vector potential, ${{\bf{A}}\left( {{\bf{R}},t} \right)}$, is retarded and transverse; it describes the photon interactions between the charged particles; the photon interactions are purely transverse. The potentials in Coulomb gauge are not Lorentz invariant. However, they are in Lorentz gauge,
\begin{equation}
\begin{array}{*{20}{l}}
{\Phi \left( {{\bf{R}},t} \right) = \int {\int {{d^3}R'dt'\frac{{\rho \left( {{{\bf{R}}^\prime },t'} \right)\delta \left( {t - t' - \left| {{\bf{R}} - {{\bf{R}}^\prime }} \right|/c} \right)}}{{\left| {{\bf{R}} - {{\bf{R}}^\prime }} \right|}}} } ;}\\
{{\bf{A}}\left( {{\bf{R}},t} \right) = \frac{1}{c}\int {\int {{d^3}R'dt'\frac{{{\bf{J}}\left( {{{\bf{R}}^\prime },t'} \right)\delta \left( {t - t' - \left| {{\bf{R}} - {{\bf{R}}^\prime }} \right|/c} \right)}}{{\left| {{\bf{R}} - {{\bf{R}}^\prime }} \right|}}} } .}
\end{array}
\label{equ14} 
\end{equation}
Here both potentials are retarded. In all gauges the relations between the true fields and the potentials are the same, viz.
\begin{equation}
\begin{array}{l}
{\bf{E}}\left( {{\bf{R}},{\bf{t}}} \right) =  - \nabla \Phi \left( {{\bf{R}},{\bf{t}}} \right) - \frac{{\bf{1}}}{{\bf{c}}}\frac{{\partial {\bf{A}}\left( {{\bf{R}},{\bf{t}}} \right)}}{{\partial {\bf{t}}}};\\
{\bf{B}}\left( {{\bf{R}},{\bf{t}}} \right) = \nabla  \times {\bf{A}}\left( {{\bf{R}},{\bf{t}}} \right).
\end{array}
\label{equ15}
\end{equation}

When retardation effects are neglected ($c \to \infty $) both gauges coincide and the only potential that remains is the instantaneous scalar potential.
\section{Background material for 2D systems\label{BG2D}}
\subsection{ Fourier transforms of a special function}
The different versions of the Fourier transform of the function $F\left( {\bf{R}} \right) = 1/R$ appear many times in the derivations so we compile the results in this subsection. This function is the Coulomb potential between two electrons, $v\left( {\bf{R}} \right) = {e^2}/R$, stripped of the charges.

The 3D Fourier transform is
\begin{equation}
F\left( {\bf{q}} \right) = \int {{d^3}R} {e^{ - i{\bf{q}} \cdot {\bf{R}}}}\frac{1}{R} = \frac{{4\pi }}{{{q^2}}};
\label{equ16}
\end{equation}
the 2D Fourier transform performed over a plane the distance $z$ from the $xy$-plane is
\begin{equation}
\begin{array}{l}
F\left( {{\bf{k}};z} \right) = \int {{d^2}r} {e^{ - i{\bf{k}} \cdot {\bf{r}}}}F\left( {{\bf{r}};z} \right) \\
= \int {{d^2}r} {e^{ - i{\bf{k}} \cdot {\bf{r}}}}\frac{1}{{\sqrt {{r^2} + {z^2}} }} = \frac{{2\pi }}{k}{e^{ - k\left| z \right|}};
\end{array}
\label{equ17}
\end{equation}
the 2D Fourier transform performed over the $xy$-plane is
\begin{equation}
\begin{array}{l}
F\left( {\bf{k}} \right) = \int {{d^2}r} {e^{ - i{\bf{k}} \cdot {\bf{r}}}}F\left( {\bf{r}} \right)  \\
=\int {{d^2}r} {e^{ - i{\bf{k}} \cdot {\bf{r}}}}\frac{1}{r} = \frac{{2\pi }}{k}.
\end{array}
\label{equ18}
\end{equation}
These relations appear in expressions involving the scalar potential in Coulomb gauge and when retardation effects are negligible. When retardation is important the function has a temporal dependence, $F\left( {{\bf{R}},t} \right) = \delta \left( {t - R/c} \right)/R$. Then the following Fourier transforms are important: The 3D Fourier transform,
\begin{equation}
\begin{array}{l}
F\left( {{\bf{q}},\omega } \right) = \int {{d^3}R\int\limits_{ - \infty }^\infty  {dt{e^{ - i\left( {{\bf{q}} \cdot {\bf{R}} - \omega t} \right)}}\frac{{\delta \left( {t - R/c} \right)}}{R}} }  \\
 =\frac{{4\pi }}{{{q^2}}}\frac{1}{{1 - {{\left( {\omega /cq} \right)}^2}}};
\end{array}
\label{equ19}
\end{equation}
the 2D Fourier transform performed over a plane the distance $z$ from the $xy$-plane,
\begin{equation}
\begin{array}{l}
F\left( {{\bf{k}};z,\omega } \right) = \int {{d^2}r} {e^{ - i{\bf{k}} \cdot {\bf{r}}}}F\left( {{\bf{r}};z,t} \right) \\
 =\int {{d^2}r\int\limits_{ - \infty }^\infty  {dt{e^{ - i\left( {{\bf{k}} \cdot {\bf{r}} - \omega t} \right)}}\frac{{\delta \left( {t - \sqrt {{r^2} + {z^2}} /c} \right)}}{{\sqrt {{r^2} + {z^2}} }}} }  \\
 =\frac{{2\pi {e^{ - k{\gamma ^{\left( 0 \right)}}\left( {k,\omega } \right)\left| z \right|}}}}{{k{\gamma ^{\left( 0 \right)}}\left( {k,\omega } \right)}};
\end{array}
\label{equ20}
\end{equation}
the 2D Fourier transform performed over the $xy$-plane,
\begin{equation}
\begin{array}{l}
F\left( {{\bf{k}},\omega } \right) = \int {{d^2}r} {e^{ - i{\bf{k}} \cdot {\bf{r}}}}F\left( {{\bf{r}},t} \right)  \\
=\int {{d^2}r\int\limits_{ - \infty }^\infty  {dt{e^{ - i\left( {{\bf{k}} \cdot {\bf{r}} - \omega t} \right)}}\frac{{\delta \left( {t - r/c} \right)}}{r}} }  \\
 =\frac{{2\pi }}{{k{\gamma ^{\left( 0 \right)}}\left( {k,\omega } \right)}},
\end{array}
\label{equ21}
\end{equation}
where in the last two equations
\begin{equation}
{\gamma ^{\left( 0 \right)}}\left( {k,\omega } \right) = \sqrt {1 - {{\left( {\omega /ck} \right)}^2}}. 
\label{equ22}
\end{equation}

We will need one further relation, viz.
\begin{equation}
\begin{array}{l}
\frac{{\partial F\left( {{\bf{k}};z,\omega } \right)}}{{\partial z}} =  - k{\gamma ^{\left( 0 \right)}}\left( {k,\omega } \right)\frac{{\left| z \right|}}{z}\frac{{2\pi {e^{ - k{\gamma ^{\left( 0 \right)}}\left( {k,\omega } \right)\left| z \right|}}}}{{k{\gamma ^{\left( 0 \right)}}\left( {k,\omega } \right)}}\\
\quad \quad \quad \quad  =  - \frac{{\left| z \right|}}{z}2\pi {e^{ - k{\gamma ^{\left( 0 \right)}}\left( {k,\omega } \right)\left| z \right|}}.
\end{array}
\label{equ23}
\end{equation}

\subsection{Method of  images\label{Images}}
When dealing with a 2D layer above a substrate it is fruitful to be able to find the image charge- and current-densities produced in the substrate. We start from a given time dependent charge- and current-density in the 2D layer, calculate the potentials and from these the $\bf E$- and  $\bf B$-fields. We make an ansatz for the image densities and determine these by using the standard boundary conditions.  Here, we use the Lorentz gauge in which the scalar and vector potentials are given in (\ref{equ14}).
Both are convolution integrals of the type 
\begin{equation}
F({\bf{R}},t) = \int {\int {{d^3}R'dt'G\left( {{\bf{R}}',t'} \right)H\left( {{\bf{R}} - {\bf{R}}',t - t'} \right)} },
\label{equ24}
\end{equation}
which has the nice property that $F({\bf{q}},\omega ) = G({\bf{q}},\omega )H({\bf{q}},\omega )$.
In the case of the scalar potential these functions are $G\left( {{\bf{R}},t} \right) = \rho \left( {{\bf{R}},t} \right)$ and $H\left( {{\bf{R}},t} \right) = \delta \left( {t - R/c} \right)/R$, respectively. For the vector potential $G\left( {{\bf{R}},t} \right) = {\bf{J}}\left( {{\bf{R}},t} \right)$ and $H\left( {{\bf{R}},t} \right) = \delta \left( {t - R/c} \right)/R$, respectively.
Thus using ({\ref{equ19}}) we find the Fourier transforms of the potentials are
\begin{equation}
\begin{array}{*{20}{l}}
{\Phi ({\bf{q}},\omega ) = \frac{{4\pi \rho ({\bf{q}},\omega )}}{{{q^2}}}\frac{1}{{1 - {{\left( {\omega /cq} \right)}^2}}};}\\
{{\bf{A}}({\bf{q}},\omega ) = \frac{{4\pi {\bf{J}}({\bf{q}},\omega )}}{{c{q^2}}}\frac{1}{{1 - {{\left( {\omega /cq} \right)}^2}}}.}
\end{array}
\label{equ25}
\end{equation}
The electric field in terms of the potentials was given in (\ref{equ15}).
This leads to the following relation between the Fourier transforms
\begin{equation}
{\bf{E}}\left( {{\bf{q}},\omega } \right) =  - i{\bf{q}}\Phi \left( {{\bf{q}},\omega } \right) + \frac{{i\omega }}{c}{\bf{A}}\left( {{\bf{q}},\omega } \right).
\label{equ26}
\end{equation}
Thus,
\begin{equation}
{\bf{E}}\left( {{\bf{q}},\omega } \right) =  - i{\bf{q}}\frac{{4\pi \rho \left( {{\bf{q}},\omega } \right)}}{{\left[ {{q^2} - {{\left( {\omega /c} \right)}^2}} \right]}} + \frac{{i\omega }}{{{c^2}}}\frac{{4\pi {\bf{J}}\left( {{\bf{q}},\omega } \right)}}{{\left[ {{q^2} - {{\left( {\omega /c} \right)}^2}} \right]}}.
\label{equ27}
\end{equation}
Now, in our system the charge- and current-densities are surface densities,
\begin{equation}
\begin{array}{*{20}{l}}
{\rho \left( {{\bf{R}},t} \right) = {\rho _s}\left( {{\bf{r}},t} \right)\delta \left( z \right);}\\
{{\bf{J}}\left( {{\bf{R}},t} \right) = {\bf{K}}\left( {{\bf{r}},t} \right)\delta \left( z \right),}
\end{array}
\label{equ28}
\end{equation}
and
\begin{equation}
\begin{array}{*{20}{l}}
{\rho \left( {{\bf{k}};z,\omega } \right) = {\rho _s}\left( {{\bf{k}},\omega } \right)\delta \left( z \right);}\\
{{\bf{J}}\left( {{\bf{k}};z,\omega } \right) = {\bf{K}}\left( {{\bf{k}},\omega } \right)\delta \left( z \right),}
\end{array}
\label{equ29}
\end{equation}
which means that
\begin{equation}
\begin{array}{l}
\rho \left( {{\bf{q}},\omega } \right) = {\rho _s} \left( {{\bf{k}},\omega } \right);\\
{\bf{J}}\left( {{\bf{q}},\omega } \right) = {\bf{K}}\left( {{\bf{k}},\omega } \right).
\end{array}
\label{equ30}
\end{equation}
Thus, we have for the electric field from surface charge- and current-densities, confined to the $xy$-plane
\begin{equation}
{\bf{E}}\left( {{\bf{q}},\omega } \right) =  - i{\bf{q}}\frac{{4\pi {\rho _s} \left( {{\bf{k}},\omega } \right)}}{{\left[ {{q^2} - {{\left( {\omega /c} \right)}^2}} \right]}} + \frac{{i\omega }}{{{c^2}}}\frac{{4\pi {\bf{K}}\left( {{\bf{k}},\omega } \right)}}{{\left[ {{q^2} - {{\left( {\omega /c} \right)}^2}} \right]}},
\label{equ31}
\end{equation}
in vacuum.   If the source densities are embedded in a medium with dielectric function ${{\tilde \varepsilon }_m}\left( \omega  \right)$ the electric field is
\begin{equation}
\begin{array}{l}
{\bf{E}}\left( {{\bf{q}},\omega } \right) =  - i{\bf{q}}\frac{{4\pi {\rho _s}\left( {{\bf{k}},\omega } \right)}}{{{{\tilde \varepsilon }_m}\left( \omega  \right)\left[ {{q^2} - {{\tilde \varepsilon }_m}\left( \omega  \right){{\left( {\omega /c} \right)}^2}} \right]}}\\
\quad \quad \quad \quad\quad \quad \quad \quad   + \frac{{i\omega }}{{{c^2}}}\frac{{4\pi {\bf{K}}\left( {{\bf{k}},\omega } \right)}}{{\left[ {{q^2} - {{\tilde \varepsilon }_m}\left( \omega  \right){{\left( {\omega /c} \right)}^2}} \right]}}.
\end{array}
\label{equ32}
\end{equation}

The magnetic induction is
\begin{equation}
{\bf{B}}\left( {{\bf{q}},\omega } \right) = i{\bf{q}} \times {\bf{A}}\left( {{\bf{q}},\omega } \right) = i\frac{1}{c}{\bf{q}} \times \frac{{4\pi {\bf{K}}\left( {{\bf{k}},\omega } \right)}}{{\left[ {{q^2} - {{\left( {\omega /c} \right)}^2}} \right]}},
\label{equ33}
\end{equation}
in vacuum and
\begin{equation}
{\bf{B}}\left( {{\bf{q}},\omega } \right) = i\frac{1}{c}{\bf{q}} \times \frac{{4\pi {\bf{K}}\left( {{\bf{k}},\omega } \right)}}{{\left[ {{q^2} - {{\tilde \varepsilon }_m}\left( \omega  \right){{\left( {\omega /c} \right)}^2}} \right]}},
\label{equ34}
\end{equation}
in a medium.

We may use the equation of continuity to eliminate ${{\rho _s}\left( {{\bf{k}},\omega } \right)}$ in favor of ${K_\parallel }\left( {{\bf{k}},\omega } \right)$. The equation of continuity reads
\begin{equation}
\nabla  \cdot {\bf{J}}({\bf{R}},t) + \frac{{\partial \rho \left( {{\bf{R}},t} \right)}}{{\partial t}} = 0.
\label{equ35}
\end{equation}
The relation for the Fourier transforms becomes
\begin{equation}
i{\bf{q}} \cdot {\bf{J}}({\bf{q}},\omega ) - i\omega \rho \left( {{\bf{q}},\omega } \right) = 0.
\label{equ36}
\end{equation}
In the present system with surface densities we have from (\ref{equ30})
\begin{equation}
i{\bf{q}} \cdot {\bf{K}}\left( {{\bf{k}},\omega } \right) = i\omega {\rho _s}\left( {{\bf{k}},\omega } \right),
\label{equ37}
\end{equation}
and since {\bf K} is in the plane
\begin{equation}
i{\bf{k}} \cdot {\bf{K}}\left( {{\bf{k}},\omega } \right) = i\omega {\rho _s}\left( {{\bf{k}},\omega } \right).
\label{equ38}
\end{equation}
The scalar product picks out the longitudinal part of the surface current density. Thus we have
\begin{equation}
{\rho _s}\left( {{\bf{k}},\omega } \right) = \left( {k/\omega } \right){K_\parallel }\left( {{\bf{k}},\omega } \right).
\label{equ39}
\end{equation}
We use this relation to eliminate the charge density from the field relations. From now on we assume that the 2D layer is embedded in a medium. 
Let us now choose the $x$-axis to point along $\bf k$. A general vector quantity has a component normal to the planar interfaces. We attach a subscript $\bf n$ to this component. The in-plane part of the vector has a longitudinal and a transverse part. The longitudinal is parallel and the transverse perpendicular to $\bf k$. We attach the subscripts $\parallel $ and $ \bot $, respectively, to these components. The current density has no normal component. To summarize we have
\begin{equation}
\begin{array}{l}
{{\bf{K}}_\parallel } = {K_x}\hat x;\quad {{\bf{E}}_\parallel } = {E_x}\hat x;\quad {{\bf{B}}_\parallel } = {B_x}\hat x;\\
{{\bf{K}}_ \bot } = {K_y}\hat y;\quad {{\bf{E}}_ \bot } = {E_y}\hat y;\quad {{\bf{B}}_ \bot } = {B_y}\hat y;\\
{{\bf{K}}_{\bf{n}}} = 0\hat z;\quad {{\bf{E}}_{\bf{n}}} = {E_z}\hat z;\quad {{\bf{B}}_{\bf{n}}} = {B_z}\hat z,
\end{array}
\label{equ40}
\end{equation}
We then have in the source plane
\begin{equation}
\begin{array}{*{20}{l}}
{{E_\parallel }\left( {{\bf{k}},\omega } \right) =  - i\frac{{2\pi }}{{{{\tilde \varepsilon }_m}\left( \omega  \right)}}\left( {k/\omega } \right){\gamma _m}\left( {k,\omega } \right){K_\parallel }\left( {{\bf{k}},\omega } \right);}\\
{{E_ \bot }\left( {{\bf{k}},\omega } \right) = \frac{{i\omega }}{{{{\tilde \varepsilon }_m}\left( \omega  \right){c^2}}}\frac{{2\pi }}{{k{\gamma _m}\left( {k,\omega } \right)}}{K_ \bot }\left( {{\bf{k}},\omega } \right);}\\
{{E_{\bf{n}}}\left( {{\bf{k}},\omega } \right) = 0;}\\
{{B_\parallel }\left( {{\bf{k}},\omega } \right) = 0;}\\
{{B_ \bot }\left( {{\bf{k}},\omega } \right) = 0;}\\
{{B_{\bf{n}}}\left( {{\bf{k}},\omega } \right) = i\frac{1}{c}\frac{{2\pi {K_ \bot }\left( {{\bf{k}},\omega } \right)}}{{{\gamma _m}\left( {k,\omega } \right)}}}
\end{array}
\label{equ41}
\end{equation}
where ${\gamma _m}\left( {k,\omega } \right) = \sqrt {1 - {{\tilde \varepsilon }_m}\left( \omega  \right){{\left( {\omega /ck} \right)}^2}}$.
In a plane parallel to the source plane ($xy$-plane) we have
\begin{equation}
\begin{array}{l}
{E_\parallel }\left( {{\bf{k}};z,\omega } \right) =  - i\frac{{2\pi }}{{{{\tilde \varepsilon }_m}\left( \omega  \right)}}\left( {k/\omega } \right){\gamma _m}\left( {k,\omega } \right)\\
\quad \quad \quad \quad \quad \quad \quad \quad  \times {e^{ - {\gamma _m}\left( {k,\omega } \right)k\left| z \right|}}{K_\parallel }\left( {{\bf{k}},\omega } \right);\\
{E_ \bot }\left( {{\bf{k}};z,\omega } \right) = \frac{{i\omega }}{{{c^2}}}\frac{{2\pi }}{{k{\gamma _m}\left( {k,\omega } \right)}}{e^{ - {\gamma _m}\left( {k,\omega } \right)k\left| z \right|}}{K_ \bot }\left( {{\bf{k}},\omega } \right);\\
{E_n}\left( {{\bf{k}};z,\omega } \right) = \frac{{2\pi }}{{{{\tilde \varepsilon }_m}\left( \omega  \right)}}\frac{z}{{\left| z \right|}}{e^{ - {\gamma _m}\left( {k,\omega } \right)k\left| z \right|}}{\rho _s}\left( {{\bf{k}},\omega } \right)\\
\quad \quad \quad \quad \quad  = \frac{{2\pi }}{{{{\tilde \varepsilon }_m}\left( \omega  \right)}}\frac{z}{{\left| z \right|}}\frac{k}{\omega }{e^{ - {\gamma _m}\left( {k,\omega } \right)k\left| z \right|}}{K_\parallel }\left( {{\bf{k}},\omega } \right);\\
{B_\parallel }\left( {{\bf{k}};z,\omega } \right) = {\gamma _m}\left( {k,\omega } \right)k\frac{z}{{\left| z \right|}}\frac{1}{c}\frac{{2\pi {K_ \bot }\left( {{\bf{k}},\omega } \right)}}{{{\gamma _m}\left( {k,\omega } \right)k}}{e^{ - {\gamma _m}\left( {k,\omega } \right)k\left| z \right|}}\\
\quad \quad \quad \quad \quad  = \frac{z}{{\left| z \right|}}\frac{1}{c}2\pi {K_ \bot }\left( {{\bf{k}},\omega } \right){e^{ - {\gamma _m}\left( {k,\omega } \right)k\left| z \right|}};\\
{B_ \bot }\left( {{\bf{k}};z,\omega } \right) =  - {\gamma _m}\left( {k,\omega } \right)k\frac{z}{{\left| z \right|}}\frac{1}{c}\frac{{2\pi {K_\parallel }\left( {{\bf{k}},\omega } \right)}}{{{\gamma _m}\left( {k,\omega } \right)k}}{e^{ - {\gamma _m}\left( {k,\omega } \right)k\left| z \right|}}\\
\quad \quad \quad \quad \quad  =  - \frac{z}{{\left| z \right|}}\frac{1}{c}2\pi {K_\parallel }\left( {{\bf{k}},\omega } \right){e^{ - {\gamma _m}\left( {k,\omega } \right)k\left| z \right|}};\\
{B_n}\left( {{\bf{k}};z,\omega } \right) = i\frac{1}{c}\frac{{2\pi {K_ \bot }\left( {{\bf{k}},\omega } \right)}}{{{\gamma _m}\left( {k,\omega } \right)}}{e^{ - {\gamma _m}\left( {k,\omega } \right)k\left| z \right|}}
\end{array}
\label{equ42}
\end{equation}
where we have used (\ref{equ39}) in the third relation.

Now, we have all we need to determine the image densities. In the spirit of image theory we assume that the fields outside the substrate can be reproduced by the fields from the actual source densities in the 2D layer (the distance $d$ from the interface) plus the fields from the mirror densities  (the distance $d$ from the interface on the opposite side ), both sources embedded in the medium with dielectric function ${{{\tilde \varepsilon }_m}\left( \omega  \right)}$. The field inside the substrate we assume can be reproduced by another mirror density at the position of the 2D layer embedded in the medium with dielectric function ${{{\tilde \varepsilon }_s}\left( \omega  \right)}$. We mark the mirror densities of the first kind by a prime and those of the second kind by a double prime. 

We start with the longitudinal current densities. We first use the boundary condition that the in plane electric field is continuous across the medium-substrate interface. This gives
\begin{equation}
\begin{array}{l}
\frac{{{\gamma _m}\left( {k,\omega } \right)}}{{{{\tilde \varepsilon }_m}\left( \omega  \right)}}{e^{ - {\gamma _m}\left( {k,\omega } \right)kd}}\left[ {{K_\parallel }\left( {{\bf{k}},\omega } \right) + {K_\parallel }'\left( {{\bf{k}},\omega } \right)} \right]\\
 = \frac{{{\gamma _s}\left( {k,\omega } \right)}}{{{{\tilde \varepsilon }_s}\left( \omega  \right)}}{e^{ - {\gamma _s}\left( {k,\omega } \right)kd}}{K_\parallel }''\left( {{\bf{k}},\omega } \right).
\end{array}
\label{equ43}
\end{equation}
Then we use the condition that the normal component of the displacement field ${\bf{\tilde D}} = \tilde \varepsilon {\bf{E}}$ is continuous across the interface. This gives
\begin{equation}
\begin{array}{l}
{e^{ - {\gamma _m}\left( {k,\omega } \right)kd}}\left[ {{K_\parallel }\left( {{\bf{k}},\omega } \right) - {K_\parallel }'\left( {{\bf{k}},\omega } \right)} \right]\\
 = {e^{ - {\gamma _s}\left( {k,\omega } \right)kd}}{K_\parallel }''\left( {{\bf{k}},\omega } \right).
\end{array}
\label{equ44}
\end{equation}
Combining these two equations results in the mirror densities
\begin{equation}
\begin{array}{l}
{K_\parallel }'\left( {{\bf{k}},\omega } \right) = \frac{{{{\tilde \varepsilon }_m}{\gamma _s} - {{\tilde \varepsilon }_s}{\gamma _m}}}{{{{\tilde \varepsilon }_m}{\gamma _s} + {{\tilde \varepsilon }_s}{\gamma _m}}}{K_\parallel }\left( {{\bf{k}},\omega } \right);\\
{K_\parallel }''\left( {{\bf{k}},\omega } \right) = \frac{{2{{\tilde \varepsilon }_s}{\gamma _m}}}{{{{\tilde \varepsilon }_m}{\gamma _s} + {{\tilde \varepsilon }_s}{\gamma _m}}}{e^{ - \left( {{\gamma _m} - {\gamma _s}} \right)kd}}{K_\parallel }\left( {{\bf{k}},\omega } \right).
\end{array}
\label{equ45}
\end{equation}

Now, we continue with the transverse current densities. We first use the boundary condition that the in plane transverse electric field is continuous across the interface. This gives
\begin{equation}
\begin{array}{*{20}{l}}
{\frac{1}{{{\gamma _m}\left( {k,\omega } \right)}}{e^{ - {\gamma _m}\left( {k,\omega } \right)kd}}\left[ {{K_ \bot }\left( {{\bf{k}},\omega } \right) + {K_ \bot }^\prime \left( {{\bf{k}},\omega } \right)} \right]}\\
{ = \frac{1}{{{\gamma _s}\left( {k,\omega } \right)}}{e^{ - {\gamma _s}\left( {k,\omega } \right)kd}}{K_ \bot }^{\prime \prime }\left( {{\bf{k}},\omega } \right).}
\end{array}
\label{equ46}
\end{equation}
Then we use the condition that the in plane longitudinal component of the ${\bf{ B}}$-field is continuous across the interface. This gives
\begin{equation}
\begin{array}{*{20}{l}}
{{e^{ - {\gamma _m}\left( {k,\omega } \right)kd}}\left[ {{K_ \bot }\left( {{\bf{k}},\omega } \right) - {K_ \bot }^\prime \left( {{\bf{k}},\omega } \right)} \right]}\\
{ = {e^{ - {\gamma _s}\left( {k,\omega } \right)kd}}{K_ \bot }^{\prime \prime }\left( {{\bf{k}},\omega } \right).}
\end{array}
\label{equ47}
\end{equation}
These two equations result in the mirror densities
\begin{equation}
\begin{array}{*{20}{l}}
{{K_ \bot }^\prime \left( {{\bf{k}},\omega } \right) = \frac{{{\gamma _m} - {\gamma _s}}}{{{\gamma _m} + {\gamma _s}}}{K_ \bot }\left( {{\bf{k}},\omega } \right);}\\
{{K_ \bot }^{\prime \prime }\left( {{\bf{k}},\omega } \right) = \frac{{2{\gamma _s}}}{{{\gamma _s} + {\gamma _m}}}{e^{ - \left( {{\gamma _m} - {\gamma _s}} \right)kd}}{K_ \bot }\left( {{\bf{k}},\omega } \right).}
\end{array}
\label{equ48}
\end{equation}
In later sections we will need the single primed mirror densities in (\ref{equ45}) and (\ref{equ48}).

\subsection{Dielectric screening in a 2D system}
The relation between the dielectric function, $\tilde {\varepsilon} $, and dynamical conductivity,  $\tilde{\sigma} $, is different in 2D and 3D. The relations are
\begin{equation}
\begin{array}{l}
{\tilde {\varepsilon}  ^{{\rm{3D}}}}\left( {{\bf{q}},\omega } \right) = 1 + {\tilde {\alpha}  ^{{\rm{3D}}}}\left( {{\bf{q}},\omega } \right) = 1 + 4\pi i{\tilde {\sigma}  ^{{\rm{3D}}}}\left( {{\bf{q}},\omega } \right)/\omega ;\\
{\tilde {\varepsilon}  ^{{\rm{2D}}}}\left( {{\bf{k}},\omega } \right) = 1 + {\tilde {\alpha}  ^{{\rm{2D}}}}\left( {{\bf{k}},\omega } \right) = 1 + 2\pi i{\tilde {\sigma}  ^{{\rm{2D}}}}\left( {{\bf{k}},\omega } \right)k/\omega,
\end{array}
\label{equ49}
\end{equation}
where $\tilde {\alpha}  $ is the polarizability. The relations between the induced current densities and the electric field are
\begin{equation}
\begin{array}{l}
{\bf{J}}\left( {{\bf{q}},\omega } \right) = {\tilde {\sigma}  ^{{\rm{3D}}}}\left( {{\bf{q}},\omega } \right){\bf{E}}\left( {{\bf{q}},\omega } \right);\\
{\bf{K}}\left( {{\bf{k}},\omega } \right) = {\tilde {\sigma}  ^{{\rm{2D}}}}\left( {{\bf{k}},\omega } \right){\bf{E}}\left( {{\bf{k}},\omega } \right).
\end{array}
\label{equ50}
\end{equation}
Note that both bound and conduction carriers contribute to the polarizabilities and conductivities.
It is sometimes convenient to introduce another correlation function, ${\chi}$, the polarization bubble (in the language of Feynman diagrams), or lowest order contribution to the density-density correlation function,
\begin{equation}
\begin{array}{*{20}{l}}
{{{\tilde \alpha }^{{\rm{3D}}}}\left( {{\bf{q}},\omega } \right) =  - {v^{{\rm{3D}}}}\left( {\bf{q}} \right){\chi ^{{\rm{3D}}}}\left( {{\bf{q}},\omega } \right) =  - 4\pi {e^2}{\chi ^{{\rm{3D}}}}\left( {{\bf{q}},\omega } \right)/{q^2};}\\
{{{\tilde \alpha }^{{\rm{2D}}}}\left( {{\bf{k}},\omega } \right) =  - {v^{{\rm{2D}}}}\left( {\bf{k}} \right){\chi ^{{\rm{2D}}}}\left( {{\bf{k}},\omega } \right) =  - 2\pi {e^2}{\chi ^{{\rm{2D}}}}\left( {{\bf{k}},\omega } \right)/k.}
\end{array}
\label{equ51}
\end{equation}

We now give the analytical expression for the 2D dielectric function of graphene and of a 2D electron gas. Let us first begin with an undoped graphene sheet. In a general point, $z$, in the complex frequency plane, away from the real axis the density-density correlation function is\cite{Guinea}
\begin{equation}
{\chi ^{{\rm{2D}}}}\left( {{\bf{k}},z} \right) =  - \frac{g}{{16\hbar }}\frac{{{k^2}}}{{\sqrt {{v^2}{k^2} - {z^2}} }},
\label{equ52}
\end{equation}
where $v$ is the carrier velocity which is a constant in graphene ($E =  \pm \hbar vk$), and $g$ represents the degeneracy parameter with the value of 4 (a factor of 2 for spin and a factor of 2 for the cone degeneracy.) In our numerical calculations\cite{Ser2,Ser3,Ser4,Ser5,Ser6} we use the value\,\cite{Wun}  $8.73723 \times {10^5}$ m/s for $v$. In the present work we do not give any numerical results.

When the graphene sheet is doped the dielectric function becomes much more complicated. However it has been derived by several groups\cite{Wun,Ser3,Sarma,KliMosSer}. 
In the two next equations we use dimension-less variables: $x = k/2{k_F}$; $y = \hbar \omega {\rm{ }}/2{E_F}$; $\tilde z = \hbar z/2{E_F}$.  
 The density-density correlation function in a general point in the complex frequency plane, $z$, away from the real axis is\,\cite{Ser3}
\begin{equation}
\begin{array}{*{20}{l}}
{{\chi ^{{\rm{2D}}}}\left( {{\bf{k}},z} \right) =  - {D_0}\left\{ {1 + \frac{{{x^2}}}{{4\sqrt {{x^2} - {{\tilde z}^2}} }}\left[ {\pi  - f\left( {x,\tilde z} \right)} \right]} \right\};}\\
{f\left( {x,\tilde z} \right) = {\rm{asin}}\left( {\frac{{1 - \tilde z}}{x}} \right) + {\rm{asin}}\left( {\frac{{1 + \tilde z}}{x}} \right)}\\
{\qquad \qquad  - \frac{{\tilde z - 1}}{x}\sqrt {1 - {{\left( {\frac{{\tilde z - 1}}{x}} \right)}^2}}  + \frac{{\tilde z + 1}}{x}\sqrt {1 - {{\left( {\frac{{\tilde z + 1}}{x}} \right)}^2}} ,}
\end{array}
\label{equ53}
\end{equation}
where ${D_0} = \sqrt {gn/\pi {\hbar ^2}{v^2}} $ is the density of states at the fermi level and $n$ is the doping concentration. 
The same result holds for excess of electrons and excess of holes. 
In the calculations one needs the function at the imaginary frequency axis. We have derived a very useful analytical expression valid along the imaginary axis; an expression in terms of real valued functions of real valued variables:\,\cite{Ser3,Ser4}
\begin{equation}
\begin{array}{*{20}{l}}
{{\chi ^{{\rm{2D}}}}\left( {{\bf{k}},i\omega } \right)}\\
{\quad \quad \quad \quad  =  - {D_0}\left\{ {1 + \frac{{{x^2}}}{{4\sqrt {{y^2} + {x^2}} }}\left[ {\pi  - g\left( {x,y} \right)} \right]} \right\};}\\
{\quad g(x,y) = {\rm{atan}}\left[ {h(x,y)k\left( {x,y} \right)} \right] + l\left( {x,y} \right);}\\
{\quad h\left( {x,y} \right) = \frac{{2{{\left\{ {{{\left[ {{x^2}\left( {{y^2} - 1} \right) + {{\left( {{y^2} + 1} \right)}^2}} \right]}^2} + {{\left( {2y{x^2}} \right)}^2}} \right\}}^{1/4}}}}{{\sqrt {{{\left( {{x^2} + {y^2} - 1} \right)}^2} + {{\left( {2y} \right)}^2}}  - \left( {{y^2} + 1} \right)}},}\\
{\quad k\left( {x,y} \right) = \sin \left\{ {\frac{1}{2}{\rm{atan}}\left[ {\frac{{2y{x^2}}}{{{x^2}\left( {{y^2} - 1} \right) + {{\left( {{y^2} + 1} \right)}^2}}}} \right]} \right\},}\\
{\quad \;l\left( {x,y} \right) = }\\
{\frac{{\sqrt { - 2{x^2}\left( {{y^2} - 1} \right) - 2\left( {{y^4} - 6{y^2} + 1} \right) + 2\left( {{y^2} + 1} \right)\sqrt {{x^4} + 2{x^2}\left( {{y^2} - 1} \right) + {{\left( {{y^2} + 1} \right)}^2}} } }}{{{x^2}}},}
\end{array}
 \label{equ54}
\end{equation}
where the arcus tangens function is taken from the branch where $ 0 \le {\rm{atan}} < \pi $. The density-density correlation function on the imaginary frequency axis has been derived before in a compact and inexplicit form (see Ref.\,\cite{Barlas} and references therein.) Here we have chosen to express it in an explicit form in terms of real valued functions of real valued variables.

The 2D polarizability of an electron gas in RPA (random phase approximation) is given by\cite{Ser7}
\begin{equation}
\begin{array}{l}
{\alpha ^{2D}}\left( {Q,iW} \right) = \\
 \quad \quad \quad \quad\frac{y}{Q}\left\{ {1 - \left[ {\sqrt {{{\left( {{Q^4} - {W^2} - {Q^2}} \right)}^2} + {{\left( {2W{Q^2}} \right)}^2}} } \right.} \right.\\
\quad \quad \quad \quad \quad \quad \quad \quad \quad  + \left. {{{\left. {{{\left( {{Q^4} - {W^2} - {Q^2}} \right)}^2}} \right]}^{1/2}}/{Q^2}} \right\},
\end{array}
\label{equ55}
\end{equation}
where
\begin{equation}
\begin{array}{l}
y = \frac{{m{e^2}}}{{{\hbar ^2}{k_F}}};\;W = \frac{{\hbar \omega }}{{4{E_F}}};\;Q = \frac{k}{{2{E_F}}};\\
{k_F} = \sqrt {2\pi {n^{2D}}} ;\;{E_F} = \frac{{{\hbar ^2}k_F^2}}{{2m}}.
\end{array}
\label{equ56}
\end{equation}

Now we have presented all background material needed for the actual derivations. 

\section{Non-retarded treatment, van der Waals interactions\label{NonRet}}
In the non-retarded treatment many terms in MEs drop out and the derivations become much simpler. There are no photons, only Coulomb interactions. The dispersion interactions treated here are called van der Waals interactions or sometimes non-retarded Casimir interactions.
We begin in section \ref{22Dnr} with two parallel 2D sheets and use different methods to derive the results. In section \ref{2DSnr} we treat a 2D sheet above a substrate.
\subsection{Two 2D films in non-retarded treatment\label{22Dnr}}
\begin{figure}
\begin{center}
\includegraphics[width=3.0cm]{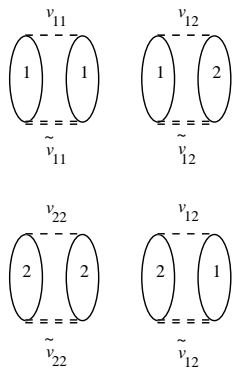}
\vspace*{8pt}
\caption{Feynman diagrams (from \cite{SerBjo}) for the correlation energy in the two 2D sheet system. The ellipses represent polarization bubbles and the dashed lines the interactions indicated in the figure. The numbers 1 and 2 refer to which sheet the carrier belongs to. See\,\cite{SerBjo} for details.}
\label{figu1}
\end{center}
\end{figure}

\subsubsection{Many-body approach in non-retarded treatment\label{Mbnr}}
For a system of parallel 2D sheets the derivation in the language of many-body theory becomes very simple, especially when retardation effects can be neglected. In the present two-sheet system the interaction energy is nothing but the inter-sheet correlation energy\,\cite{SerBjo}. Using diagrammatic perturbation theory the Feynman diagrams representing the correlation energy are given in Fig.\,\ref{figu1}. To get the inter-sheet contribution we can either subtract the intra-sheet part or subtract the total result when the separation between the sheets goes to infinity (at that limit only the intra-sheet contribution remains.) To save space we assume that the two sheets are identical. It is straight forward to extend the treatment to different sheets. Each ellipse represents a polarization bubble,  $\chi \left( {{\bf{k}},\omega } \right)$, and the number indicates in which sheet the process occurs. Each of the four diagrams represents an infinite series of diagrams. Since the sheets are identical the diagrams in the second row give  contributions identical to the diagrams in the first row. We can use the first row diagrams and multiply the result with a factor of two. The interaction energy per unit area can be written as\cite{Ser6}
\begin{equation}
\begin{array}{l}
{E_c}\left( d \right) = \hbar \int {\frac{{{d^2}k}}{{{{\left( {2\pi } \right)}^2}}}} \int\limits_0^\infty  {\frac{{d\omega }}{{2\pi }}\int\limits_0^1 {d\lambda \frac{1}{\lambda }} 2\left[ {Diag1\left( {k,i\omega ;\lambda } \right)} \right.} \\
\left. {\quad \quad \quad \quad \quad \quad \quad \quad \quad \quad \quad \quad  + Diag2\left( {k,i\omega ;\lambda } \right)} \right],
\end{array}
\label{equ57}
\end{equation}
where $\lambda$ is the coupling constant and the factor of two has been inserted. 

Now ${v_{11}} = {v^{2D}}\left( {{\bf{k}},\omega } \right)$ and ${v_{12}} = \exp \left( { - kd} \right){v^{2D}}\left( {{\bf{k}},\omega } \right)$, respectively. Note that with our assumptions ${v_{22}} ={v_{11}}$ and ${v_{21}} ={v_{12}}$.  The distance between the sheets is $d$ and ${v^{2D}}$ is the 2D Coulomb interaction, ${v^{2D}}\left( {\bf{r}} \right) = {e^2}/r$. The exponential factor in the second potential is the result from taking the 2D Fourier transform of the Coulomb potential in a plane the distance $d$ from the center of the potential, see (\ref{equ17}). The interaction lines with double bars represent a series of terms, with zero, one, two $ \ldots $ number of polarization bubbles. This can be expressed as
\begin{eqnarray}
\nonumber {{\tilde v}_{11}}\left( {{\bf{k}},\omega } \right) &=& {v^{2D}}\left( k  \right) + {v^{2D}}\left( k  \right)\chi \left( {{\bf{k}},\omega } \right){{\tilde v}_{11}}\left( {{\bf{k}},\omega } \right) \\
\nonumber&&+ \exp \left( { - kd} \right){v^{2D}}\left( k  \right)\chi \left( {{\bf{k}},\omega } \right){{\tilde v}_{12}}\left( {{\bf{k}},\omega } \right)\\
\nonumber {{\tilde v}_{12}}\left( {{\bf{k}},\omega } \right) &= &\exp \left( { - kd} \right){v^{2D}}\left( k  \right) + {v^{2D}}\left( k  \right)\chi \left( {{\bf{k}},\omega } \right){{\tilde v}_{12}}\left( {{\bf{k}},\omega } \right)\\
& &+ \exp \left( { - kd} \right){v^{2D}}\left( k  \right)\chi \left( {{\bf{k}},\omega } \right){{\tilde v}_{11}}\left( {{\bf{k}},\omega } \right),
\label{equ58}
\end{eqnarray}
where we have closed the two infinite series. This system of equations can be solved and the result is
%
\begin{eqnarray}
\nonumber {{\tilde v}_{11}}\left( {{\bf{k}},\omega } \right) = \frac{{{v^{2D}}\left( k \right)\left[ {1 + \tilde {\alpha}  \left( {{\bf{k}},\omega } \right)\left( {1 - \exp \left( { - 2kd} \right)} \right)} \right]}}{{{{\left[ {1 + \tilde {\alpha}  \left( {{\bf{k}},\omega } \right)} \right]}^2} - \exp \left( { - 2kd} \right){\tilde {\alpha}  ^2}\left( {{\bf{k}},\omega } \right)}};\\
{{\tilde v}_{12}}\left( {{\bf{k}},\omega } \right) = \frac{{{v^{2D}}\left( k \right)\exp \left( { - kd} \right)}}{{{{\left[ {1 + \tilde {\alpha}  \left( {{\bf{k}},\omega } \right)} \right]}^2} - \exp \left( { - 2kd} \right){\tilde {\alpha}  ^2}\left( {{\bf{k}},\omega } \right)}},
\label{equ59}
\end{eqnarray}
where we have used the relation $\tilde \alpha \left( {{\bf{k}},\omega } \right) =- {v^{{\rm{2D}}}}\left( k \right)\chi \left( {{\bf{k}},\omega } \right)$ in accordance with (\ref{equ51}).
Interpreting the Feynman diagrams in figure \ref{figu1} one finds that the square brackets in (\ref{equ57}) are $\left[ {{{\tilde v}_{11}}{v_{11}}{\chi ^2} + {{\tilde v}_{12}}{v_{12}}{\chi ^2} - \left( {{v^{2D}}/\varepsilon } \right){v^{2D}}{\chi ^2}} \right]$, where the last term comes from the subtraction of the intra-layer correlation energy. Each factor of ${e^2}$ appearing implicitly in the expression should be multiplied by the coupling constant. Performing the integration over coupling constant gives
\begin{equation}
{E_c}\left( d \right) = \hbar \int {\frac{{{d^2}k}}{{{{\left( {2\pi } \right)}^2}}}\int\limits_0^\infty  {\frac{{d\omega }}{{2\pi }}\ln \left\{ {1 - {e^{ - 2kd}}{{\left[ {\frac{{\tilde {\alpha}  \left( {k,i\omega } \right)}}{{1 + \tilde {\alpha}  \left( {k,i\omega } \right)}}} \right]}^2}} \right\}} }.
\label{equ60}
\end{equation}
Here for simplicity we assumed that the two 2D layers were identical. If they are not the result is
\begin{equation}
\begin{array}{l}
{E_c}\left( d \right) = \hbar \int {\frac{{{d^2}k}}{{{{\left( {2\pi } \right)}^2}}}\int\limits_0^\infty  {\frac{{d\omega }}{{2\pi }}} } \\
\quad\quad\quad  \times \ln \left\{ {1 - {e^{ - 2kd}}\left[ {\frac{{{{\tilde \alpha }_1}\left( {k,i\omega } \right)}}{{1 + {{\tilde \alpha }_1}\left( {k,i\omega } \right)}}} \right]\left[ {\frac{{{{\tilde \alpha }_2}\left( {k,i\omega } \right)}}{{1 + {{\tilde \alpha }_2}\left( {k,i\omega } \right)}}} \right]} \right\}.
\end{array}
\label{equ61}
\end{equation}
This is the result from diagrammatic perturbation theory within the random phase approximation (RPA). In next section we derive the same thing using normal modes.

\subsubsection{Coupled induced sources in non-retarded treatment\label{Cisnr}}
We will now go through one possible way to find the normal modes and their zero-point energies. Let us assume that we have an induced charge distribution, $\rho_1 \left( {{\bf{k}},\omega } \right)$, in sheet 1. This gives rise to the potential $v\left( {{\bf{k}},\omega } \right) = {{v^{2D} \left( k \right)\rho _1 \left( {{\bf{k}},\omega } \right)} }$ and ${{\exp \left( { - kd} \right)v^{2D} \left( k \right)\rho _1 \left( {{\bf{k}},\omega } \right)}}$
 in sheets 1 and 2, respectively. The resulting potential in sheet 2 after screening by the carriers is $\exp \left( { - kd} \right){v^{2D}}\left( k \right){\rho _1}\left( {{\bf{k}},\omega } \right)/\left[ {1 + \tilde {\alpha}_2 \left( {{\bf{k}},\omega } \right)} \right]$, which gives rise to an induced charge distribution in sheet 2, 
\begin{equation}
\rho _2 \left( {{\bf{k}},\omega } \right) = \chi_2 \left( {{\bf{k}},\omega } \right)e^{ - kd} v^{2D} \left( k \right)\frac{{\rho _1 \left( {{\bf{k}},\omega } \right)}}{{\ \left[ {1 + \tilde {\alpha}_2  \left( {{\bf{k}},\omega } \right)} \right]}}.
\label{equ62}
\end{equation}
In complete analogy, this charge distribution in sheet 2 gives rise to a charge distribution in sheet 1,
\begin{equation}
\rho _1 \left( {{\bf{k}},\omega } \right) = \chi_1 \left( {{\bf{k}},\omega } \right)e^{ - kd} v^{2D} \left( k \right)\frac{{\rho _2 \left( {{\bf{k}},\omega } \right)}}{{ \left[ {1 + \tilde {\alpha}_2  \left( {{\bf{k}},\omega } \right)} \right]}}.
\label{equ63}
\end{equation}
To find the condition for self-sustained fields, normal modes, we let this induced charge density in sheet 1 be the charge density we started from. This leads to
\begin{equation}
1 - {e^{ - 2kd}}\left[ {\frac{{{{\tilde \alpha }_1}\left( {{\bf{k}},\omega } \right)}}{{1 + {{\tilde \alpha }_1}\left( {{\bf{k}},\omega } \right)}}} \right]\left[ {\frac{{{{\tilde \alpha }_2}\left( {{\bf{k}},\omega } \right)}}{{1 + {{\tilde \alpha }_2}\left( {{\bf{k}},\omega } \right)}}} \right] = 0.
\label{equ64}
\end{equation}
The left hand side of this equation (${f_{\bf{k}}}\left( \omega  \right)$) is exactly the argument of the logarithm in \,(\ref{equ61}). Inserting this function into \,(\ref{equ2}) we obtain exactly the same energy as in \,(\ref{equ61}). That result was derived using many-body theory and in that derivation there is no talk about zero-point energies at all. 

\subsubsection{Maxwells equations and boundary conditions in non-retarded treatment\label{MEnr22D}}
Here we solve MEs in the three regions separated by the two 2D layers. We treat the induced charge and current densities in the 2D layers as external to our system. Since we have vacuum in between the layers our system is vacuum and the $\bf D$ and $\bf H$ fields are equal to the $\bf E$ and $\bf B$ fields, respectively. It is enough to study the $\bf E$-field in the non-retarded treatment. We follow the procedure in \cite{Ser1} and make the following ansatz
\begin{equation}
{\bf{E}} = \left\{ \begin{array}{l}
\left( {{{\bf{E}}^{{\rm{BR}}}}{e^{kz}} + {{\bf{E}}^{{\rm{BL}}}}{e^{ - k\left( {z + d} \right)}}} \right){e^{i\left( {kx - \omega t} \right)}},\; - d \le z \le 0\\
{{\bf{E}}^{\rm{R}}}{e^{ - kz}}{e^{i\left( {kx - \omega t} \right)}},\;0 \le z\\
{{\bf{E}}^{\rm{L}}}{e^{k\left( {z + d} \right)}}{e^{i\left( {kx - \omega t} \right)}},\;z \le  - d.
\end{array} \right.
\label{equ65}
\end{equation}
The two MEs involving the $\bf E$-field, $\nabla  \cdot {\bf{E}} = 0$, and $\nabla  \times {\bf{E}} = 0$, both give
\begin{equation}
\begin{array}{l}
{{E}}_z^{{\rm{BR}}} =  - i{{E}}_x^{{\rm{BR}}};\\
{{E}}_z^{{\rm{BL}}} = i{{E}}_x^{{\rm{BL}}};\\
{{E}}_z^{\rm{R}} = i{{E}}_x^{\rm{R}};\\
{{E}}_z^{\rm{L}} =  - i{{E}}_x^{\rm{L}}.
\end{array}
\label{equ66}
\end{equation}
The boundary conditions are that the $x$-components of the fields are continuous across the two interfaces and there is a jump in the $z$-component equal to $4\pi {\rho _s}$. Thus we have
\begin{equation}
\begin{array}{l}
{{E}}_x^{{\rm{BR}}} + {e^{ - kd}}{{E}}_x^{{\rm{BL}}} = {{E}}_x^{\rm{R}};\\
{e^{ - kd}}{{E}}_x^{{\rm{BR}}} + {{E}}_x^{{\rm{BL}}} = {{E}}_x^{\rm{L}},
\end{array}
\label{equ67}
\end{equation}
or on matrix form
\begin{equation}
\left( {\begin{array}{*{20}{c}}
1&{{e^{ - kd}}}\\
{{e^{ - kd}}}&1
\end{array}} \right) \cdot \left( {\begin{array}{*{20}{c}}
{E_x^{{\rm{BR}}}}\\
{E_x^{{\rm{BL}}}}
\end{array}} \right) = \left( {\begin{array}{*{20}{c}}
{E_x^{\rm{R}}}\\
{E_x^{\rm{L}}}
\end{array}} \right),
\label{equ68}
\end{equation}
and
\begin{equation}
\begin{array}{l}
{{E}}_z^{{\rm{BR}}} + {e^{ - kd}}{{E}}_z^{{\rm{BL}}} + 4\pi \rho _s^R = {{E}}_z^{\rm{R}};\\
{e^{ - kd}}{{E}}_z^{{\rm{BR}}} + {{E}}_z^{{\rm{BL}}} = {{E}}_z^{\rm{L}} + 4\pi \rho _s^L.
\end{array}
\label{equ69}
\end{equation}
Now, we eliminate the $z$-components in favor of the $x$-components using (\ref{equ66}) and use ${\rho _s} = kK/\omega  = k\tilde {\sigma}  {{{E}}_x}/\omega $ and assume for simplicity that the two layers are identical. We find
\begin{equation}
\begin{array}{l}
{{E}}_x^{{\rm{BR}}} - {e^{ - kd}}{{E}}_x^{{\rm{BL}}} =  - {{E}}_x^{\rm{R}}\left( {1 + i4\pi k\tilde {\sigma}  /\omega } \right);\\
 - {e^{ - kd}}{{E}}_x^{{\rm{BR}}} + {{E}}_x^{{\rm{BL}}} =  - {{E}}_x^{\rm{L}}\left( {1 + i4\pi k\tilde {\sigma}  /\omega } \right),
\end{array}
\label{equ70}
\end{equation}
or on matrix form
\begin{equation}
\left( {\begin{array}{*{20}{c}}
1&{ - {e^{ - kd}}}\\
{ - {e^{ - kd}}}&1
\end{array}} \right)\cdot \left( {\begin{array}{*{20}{c}}
{{{E}}_x^{{\rm{BR}}}}\\
{{{E}}_x^{{\rm{BL}}}}
\end{array}} \right) =  - \left( {2\tilde \varepsilon  - 1} \right)\left( {\begin{array}{*{20}{c}}
{{{E}}_x^{\rm{R}}}\\
{{{E}}_x^{\rm{L}}}
\end{array}} \right).
\label{equ71}
\end{equation}
Combining (\ref{equ68}) and (\ref{equ71}) we find that the condition for having normal modes is that the determinant of the matrix
\begin{equation}
\begin{array}{*{20}{l}}
{\left( {\begin{array}{*{20}{c}}
1&{ - {e^{ - kd}}}\\
{ - {e^{ - kd}}}&1
\end{array}} \right) + \left( {2\tilde \varepsilon  - 1} \right)\left( {\begin{array}{*{20}{c}}
1&{{e^{ - kd}}}\\
{{e^{ - kd}}}&1
\end{array}} \right)}\\
\begin{array}{l}
 = \left( {\begin{array}{*{20}{c}}
{2\tilde \varepsilon }&{2\left( {\tilde \varepsilon  - 1} \right){e^{ - kd}}}\\
{2\left( {\tilde \varepsilon  - 1} \right){e^{ - kd}}}&{2\tilde \varepsilon }
\end{array}} \right)\\
 = 2\left( {\begin{array}{*{20}{c}}
{1 + \tilde \alpha }&{\tilde \alpha {e^{ - kd}}}\\
{\tilde \alpha {e^{ - kd}}}&{1 + \tilde \alpha }
\end{array}} \right)
\end{array}
\end{array}
\label{equ72}
\end{equation}
vanishes. The mode condition function becomes
\begin{equation}
{f_{\bf{k}}}\left( \omega  \right) = 1 - {e^{ - 2kd}}{\left[ {\frac{{\tilde \alpha \left( {{\bf{k}},\omega } \right)}}{{1 + \tilde \alpha \left( {{\bf{k}},\omega } \right)}}} \right]^2}.
\label{equ73}
\end{equation}
We have taken as reference system the system when the two sheets are at infinite separation. This means that we have divided the mode condition function with the corresponding result when $d$ is infinite.
If the two layers are different we instead find
\begin{equation}
{f_{\bf{k}}}\left( \omega  \right) = 1 - {e^{ - 2kd}}\left[ {\frac{{{{\tilde \alpha }_1}\left( {{\bf{k}},\omega } \right)}}{{1 + {{\tilde \alpha }_1}\left( {{\bf{k}},\omega } \right)}}} \right]\left[ {\frac{{{{\tilde \alpha }_2}\left( {{\bf{k}},\omega } \right)}}{{1 + {{\tilde \alpha }_2}\left( {{\bf{k}},\omega } \right)}}} \right],
\label{equ74}
\end{equation}
which agrees with the results of the previous sections.

\subsubsection{Coupled potentials in non-retarded treatment\label{Cpnr}}
Another way to find the normal modes is to find the frequency where the potential between carriers in the same layer or in different layers diverges. The condition for modes is that the denominators in (\ref{equ59}) are equal to zero. This leads to the same mode condition function as in (\ref{equ64}) and (\ref{equ74}) when we use as reference system the system when the 2D layers are at infinite separation.

\subsection{One 2D film above a substrate in non-retarded treatment\label{2DSnr}}

\subsubsection{Coupled induced sources in non-retarded treatment}
If we have a 2D layer (like a graphene sheet) the distance $d$ above a substrate the procedure is very similar to in section \ref{Cisnr}. We start with an induced mirror charge density,  $\rho_1 \left( {{\bf{k}},\omega } \right)$, in the substrate. The induced charge density in the 2D layer is given by the expression in (\ref{equ62}) except that now the distance between the mirror charge and the 2D layer is $2d$ instead of $d$. The mirror charges can be obtained from  (\ref{equ39}) and  (\ref{equ45}) and agree with the traditional results for the mirror charges at a planar vacuum dielectric interface.\,\cite{Jackson} Equation\,(\ref{equ63}) is then replaced by 
\begin{equation}
\rho _1 \left( {{\bf{k}},\omega } \right) =  - \rho _2 \left( {{\bf{k}},\omega } \right)\frac{{\tilde {\varepsilon} _s \left( \omega  \right) - 1}}{{\tilde {\varepsilon} _s \left( \omega  \right) + 1}},
\label{equ75}
\end{equation}
and the condition for normal modes becomes
\begin{equation}
1 - e^{ - 2kd} \frac{{\tilde {\alpha}  \left( {{\bf{k}},\omega } \right)}}{{1 + \tilde {\alpha}  \left( {{\bf{k}},\omega } \right)}}\frac{{\tilde {\varepsilon} _s \left( \omega  \right) - 1}}{{\tilde {\varepsilon} _s \left( \omega  \right) + 1}} = 0,
\label{equ76}
\end{equation}
and hence the mode condition function is
\begin{equation}
{f_{\bf{k}}}\left( \omega  \right) = 1 - {e^{ - 2kd}}\frac{{\tilde \alpha \left( {{\bf{k}},\omega } \right)}}{{1 + \tilde \alpha \left( {{\bf{k}},\omega } \right)}}\frac{{{{\tilde \varepsilon }_s}\left( \omega  \right) - 1}}{{{{\tilde \varepsilon }_s}\left( \omega  \right) + 1}}.
\label{equ77}
\end{equation}

\subsubsection{Maxwells equations and boundary conditions in non-retarded treatment\label{MEnr2DS}}
In section \ref{MEnr22D} we had a geometry with two interfaces, one at $z=-d$ and one at $z=0$. There there were a 2D layer at both interfaces. We use the same geometry here but there is a 2D layer only at $z=-d$. At $z=0$ we place the vacuum-substrate interface. The derivation and equations are equal up till (\ref{equ69}). When the media are no longer vacuum  there is a jump in ${{{{\tilde D}}}_z}$ at the 2D layer instead of in ${{{E}}_z}$. Note that here there is only a 2D layer at the left interface so the normal component of the $\tilde{D}$-field is continuous at the substrate surface. The new equation is
\begin{equation}
\begin{array}{l}
{{E}}_z^{{\rm{BR}}} + {e^{ - kd}}{{E}}_z^{{\rm{BL}}} = {{\tilde \varepsilon }_s}{{E}}_z^{\rm{R}};\\
{e^{ - kd}}{{E}}_z^{{\rm{BR}}} + {{E}}_z^{{\rm{BL}}} = {{E}}_z^{\rm{L}} + 4\pi {\rho _s}.
\end{array}
\label{equ78}
\end{equation}

Now we eliminate the $z$-components in favor of the $x$-components and use ${\rho _s} = kK/\omega  = k\tilde {\sigma}  {{{E}}_x}/\omega$ so that
\begin{equation}
\begin{array}{l}
{{E}}_x^{{\rm{BR}}} - {e^{ - kd}}{{E}}_x^{{\rm{BL}}} =  - {{\tilde \varepsilon }_s}{{E}}_x^{\rm{R}};\\
 - {e^{ - kd}}{{E}}_x^{{\rm{BR}}} + {{E}}_x^{{\rm{BL}}} =  - {{E}}_x^{\rm{L}}\left( {1 + i4\pi k\tilde {\sigma}  /\omega } \right),
\end{array}
\label{equ79}
\end{equation}
or on matrix form
\begin{equation}
\left( {\begin{array}{*{20}{c}}
1&{ - {e^{ - kd}}}\\
{ - {e^{ - kd}}}&1
\end{array}} \right)\cdot \left( {\begin{array}{*{20}{c}}
{{{E}}_x^{{\rm{BR}}}}\\
{{{E}}_x^{{\rm{BL}}}}
\end{array}} \right) =  - \left( {\begin{array}{*{20}{c}}
{{{\tilde \varepsilon }_s}}&0\\
0&{2\tilde \varepsilon  - 1}
\end{array}} \right)\cdot \left( {\begin{array}{*{20}{c}}
{{{E}}_x^{\rm{R}}}\\
{{{E}}_x^{\rm{L}}}
\end{array}} \right).
\label{equ80}
\end{equation}
Combining (\ref{equ68}) and (\ref{equ80}) we find that the condition for having normal modes is that the determinant of the matrix
\begin{equation}
\begin{array}{*{20}{l}}
{\left( {\begin{array}{*{20}{c}}
1&{ - {e^{ - kd}}}\\
{ - {e^{ - kd}}}&1
\end{array}} \right) + \left( {\begin{array}{*{20}{c}}
{{{\tilde \varepsilon }_s}}&0\\
0&{2\tilde \varepsilon  - 1}
\end{array}} \right).\left( {\begin{array}{*{20}{c}}
1&{{e^{ - kd}}}\\
{{e^{ - kd}}}&1
\end{array}} \right)}\\
\begin{array}{l}
 = \left( {\begin{array}{*{20}{c}}
{{{\tilde \varepsilon }_s} + 1}&{\left( {{{\tilde \varepsilon }_s} - 1} \right){e^{ - kd}}}\\
{2\left( {\tilde \varepsilon  - 1} \right){e^{ - kd}}}&{2\tilde \varepsilon }
\end{array}} \right)\\
 = \left( {\begin{array}{*{20}{c}}
{{{\tilde \varepsilon }_s} + 1}&{\left( {{{\tilde \varepsilon }_s} - 1} \right){e^{ - kd}}}\\
{2\tilde \alpha {e^{ - kd}}}&{2\tilde \varepsilon }
\end{array}} \right)
\end{array}
\end{array}
\label{equ81}
\end{equation}
is zero. Thus, the mode condition function becomes
\begin{equation}
{f_{\bf{k}}}\left( \omega  \right) = 1 - {e^{ - 2kd}}\left[ {\frac{{\tilde \alpha \left( {{\bf{k}},\omega } \right)}}{{1 + \tilde \alpha \left( {{\bf{k}},\omega } \right)}}} \right]\left[ {\frac{{{{\tilde \varepsilon }_s}\left( \omega  \right) - 1}}{{{{\tilde \varepsilon }_s}\left( \omega  \right) + 1}}} \right],
\label{equ82}
\end{equation}
which agrees with the results of the previous sections.

\section{Retarded treatment, Casimir interactions\label{Ret}}
Here we take retardation effects fully into account. The formalism becomes much more involved. The dispersion interactions treated here are called Casimir interactions or sometimes retarded van der Waals interactions. We begin in section \ref{22Dret} with two parallel 2D sheets and use different methods to derive the results. In section \ref{2DSret} we treat a 2D sheet above a substrate.
\subsection{Two 2D films in fully retarded treatment\label{22Dret}}
\subsubsection{Many-body approach in fully retarded treatment\label{MBret}}
We do not perform a detailed derivation here, we just describe how the derivation is performed. We follow the derivation in \cite{SerBjo} and refer to that work for more details. We have chosen to work in Coulomb gauge. In this gauge one part of the interaction between the electrons is in the form of the instantaneous, longitudinal Coulomb interaction (scalar potential) and the other part is via transverse photons (vector potential). The Hamiltonian for the system is
\begin{equation}
\begin{array}{l}
H = \sum\limits_i {\frac{1}{{2m*}}} {\left[ {{{\bf{p}}_i} - \frac{{{e_i}}}{c}{\bf{A}}\left( {{{\bf{R}}_i}} \right)} \right]^2} + \frac{1}{2}\sum\limits_{ij} {\frac{{{e_i}{e_j}}}{{{R_{ij}}}}} \\
\quad  + \sum\limits_{{\bf{q}},\lambda } {\hbar {\omega _{\bf{q}}}\left( {a_{{\bf{q}}\lambda }^\dag {a_{{\bf{q}}\lambda }} + 1/2} \right)},
\end{array}
\label{equ83}
\end{equation}
where the first term is the kinetic energy that contains the interactions via the vector potential $\bf A$.  The second term represents the scalar potential interaction and the last term is the free photon Hamiltonian. The non-retarded results are obtained by letting the speed of light tend to infinity. From the Hamiltonian we see that this corresponds to neglecting the
vector-potential interaction completely. Now, in the specific system we consider here, the electrons are only free to move
in a plane. In a plane only one longitudinal and one transverse electric field can exist. It turns out that the longitudinal electric field and the $p$-polarized photons combine into one field that is longitudinal in the plane. The induced charge and
current densities from such a field will also produce a field of that kind. The $s$-polarized photons produce an electric field
that is transverse in the plane. The current induced by such a field produces a field of the same type. 

The two types of interaction involving longitudinal and transverse electric fields in the planes defined by the two sheets will not mix and the energy is given by a sum of two sets of diagrams of the type given in figure\,\ref{figu1}, one set for the longitudinal interaction and one containing only transverse interactions. 

Let us now discuss the longitudinal interaction. The coupling to the longitudinal field and $p$-polarized photons is given by the scalar potential and the ${\bf{p}} \cdot {\bf{A}}$ terms in the Hamiltonian. The resulting interaction energy per unit area is
\begin{equation}
\begin{array}{l}
E_c^{l + p}\left( d \right) = \hbar \int {\frac{{{d^2}k}}{{{{\left( {2\pi } \right)}^2}}}\int\limits_0^\infty  {\frac{{d\omega }}{{2\pi }}} } \\
\quad \quad  \times \ln \left\{ {1 - {e^{ - 2{\gamma ^{\left( 0 \right)}}\left( {k,i\omega } \right)kd}}{{\left[ {\frac{{{\gamma ^{\left( 0 \right)}}\left( {k,i\omega } \right){{\tilde \alpha }_\parallel }\left( {k,i\omega } \right)}}{{1 + {\gamma ^{\left( 0 \right)}}\left( {k,i\omega } \right){{\tilde \alpha }_\parallel }\left( {k,i\omega } \right)}}} \right]}^2}} \right\}.
\end{array}
\label{equ84}
\end{equation}
Compared with the non-retarded result for the correlation energy, we gave earlier in (\ref{equ60}), we find that the polarizability has attained a factor ${{\gamma ^{\left( 0 \right)}}\left( {k,i\omega } \right)}$ in front,
which is in agreement with the polarizability obtained by Stern\,\cite{Stern}.
Furthermore, the factor appearing in the interlayer interactions is modified with the same factor in the exponent.

The response to the $s$-polarized fields is a transverse current. It is described by the transverse conductivity, which is dominated by the contribution originating from the ${{\bf{A}}^2}$-term in the Hamiltonian. Also the ${\bf{p}} \cdot {\bf{A}}$
terms in the Hamiltonian give a contribution to the transverse conductivity, in the form of the current-current correlation function. The energy per unit area from the $s$-photon interaction is
\begin{equation}
\begin{array}{l}
E_c^s\left( d \right) = \hbar \int {\frac{{{d^2}k}}{{{{\left( {2\pi } \right)}^2}}}\int\limits_0^\infty  {\frac{{d\omega }}{{2\pi }}} } \\
\quad \quad  \times \ln \left\{ {1 - {e^{ - 2{\gamma ^{\left( 0 \right)}}\left( {k,i\omega } \right)kd}}{{\left[ {\frac{{ - {{\left( {\omega /ck} \right)}^2}{{\tilde \alpha }_ \bot }\left( {k,i\omega } \right)}}{{{\gamma ^{\left( 0 \right)}}\left( {k,i\omega } \right) + {{\left( {\omega /ck} \right)}^2}{{\tilde \alpha }_ \bot }\left( {k,i\omega } \right)}}} \right]}^2}} \right\}.
\end{array}
\label{equ85}
\end{equation}

\subsubsection{Coupled induced sources in fully retarded treatment}
We could here proceed in a way analogous to how we derived the modes in section \ref{Cisnr} by starting with a current density (longitudinal or transverse) in layer 1. This current density gives rise to an electric field in layer 2; this field induces a current density in layer 2; this current density in layer 2 gives rise to an electric field in layer 1 which causes an induced current density in layer 1. If we let this current density be the one we started with we have closed the loop and produced self-sustained fields, normal modes. When the current densities and electric fields are longitudinal the modes are of TM type and the TE modes are the results when the current densities and fields are transverse.

To vary the derivations we instead proceed in a slightly different way. We have two layers, 1 and 2. The electric field in layer 1 has contributions from both layers and the induced current density in layer 1 is the conductivity of layer 1 times the electric field in layer 1. The corresponding is true for layer 2. We begin with the TM modes. The longitudinal field in layer 1 is from (\ref{equ41}) and (\ref{equ42})

\begin{equation}
\begin{array}{l}
E_\parallel ^1\left( {{\bf{k}},\omega } \right) =  - i2\pi \left( {k/\omega } \right){\gamma ^{\left( 0 \right)}}\left( {k,\omega } \right)\\
\quad \quad \quad \quad  \times \left[ {K_\parallel ^1\left( {{\bf{k}},\omega } \right) + {e^{ - {\gamma ^{\left( 0 \right)}}\left( {k,\omega } \right)kd}}K_\parallel ^2\left( {{\bf{k}},\omega } \right)} \right]
\end{array},
\label{equ86}
\end{equation}
and the induced current density is
\begin{equation}
\begin{array}{*{20}{l}}
{K_\parallel ^1\left( {{\bf{k}},\omega } \right) = \tilde \sigma _\parallel ^1\left( {{\bf{k}},\omega } \right)E_\parallel ^1\left( {{\bf{k}},\omega } \right)}\\
\begin{array}{l}
\quad \quad \quad  =  - i\tilde \sigma _\parallel ^1\left( {{\bf{k}},\omega } \right)2\pi \left( {k/\omega } \right){\gamma ^{\left( 0 \right)}}\left( {k,\omega } \right)\\
\quad \quad \quad \quad  \times \left[ {K_\parallel ^1\left( {{\bf{k}},\omega } \right) + {e^{ - {\gamma ^{\left( 0 \right)}}\left( {k,\omega } \right)kd}}K_\parallel ^2\left( {{\bf{k}},\omega } \right)} \right]
\end{array}\\
\begin{array}{l}
\quad \quad \quad  =  - \tilde \alpha _\parallel ^1\left( {{\bf{k}},\omega } \right){\gamma ^{\left( 0 \right)}}\left( {k,\omega } \right)\\
\quad \quad \quad \quad  \times \left[ {K_\parallel ^1\left( {{\bf{k}},\omega } \right) + {e^{ - {\gamma ^{\left( 0 \right)}}\left( {k,\omega } \right)kd}}K_\parallel ^2\left( {{\bf{k}},\omega } \right)} \right].
\end{array}
\end{array}
\label{equ87}
\end{equation}
In analogy we get the result for the induced current density in layer 2 by interchanging the superscripts. Thus we have
\begin{equation}
\begin{array}{*{20}{l}}
\begin{array}{l}
K_\parallel ^1\left( {{\bf{k}},\omega } \right) =  - \tilde \alpha _\parallel ^1\left( {{\bf{k}},\omega } \right){\gamma ^{\left( 0 \right)}}\left( {k,\omega } \right)\\
\quad \quad \quad \quad  \times \left[ {K_\parallel ^1\left( {{\bf{k}},\omega } \right) + {e^{ - {\gamma ^{\left( 0 \right)}}\left( {k,\omega } \right)kd}}K_\parallel ^2\left( {{\bf{k}},\omega } \right)} \right];
\end{array}\\
\begin{array}{l}
K_\parallel ^2\left( {{\bf{k}},\omega } \right) =  - \tilde \alpha _\parallel ^2\left( {{\bf{k}},\omega } \right){\gamma ^{\left( 0 \right)}}\left( {k,\omega } \right)\\
\quad \quad \quad \quad  \times \left[ {{e^{ - {\gamma ^{\left( 0 \right)}}\left( {k,\omega } \right)kd}}K_\parallel ^1\left( {{\bf{k}},\omega } \right) + K_\parallel ^2\left( {{\bf{k}},\omega } \right)} \right],
\end{array}
\end{array}
\label{equ88}
\end{equation}
and on matrix form
\begin{equation}
\begin{array}{l}
\left( {\begin{array}{*{20}{c}}
{ - \tilde \alpha _\parallel ^1{\gamma ^{\left( 0 \right)}} - 1}&{ - {e^{ - {\gamma ^{\left( 0 \right)}}kd}}\tilde \alpha _\parallel ^1{\gamma ^{\left( 0 \right)}}}\\
{ - {e^{ - {\gamma ^{\left( 0 \right)}}kd}}\tilde \alpha _\parallel ^2{\gamma ^{\left( 0 \right)}} - 1}&{ - \tilde \alpha _\parallel ^2{\gamma ^{\left( 0 \right)}}}
\end{array}} \right) \\
 \cdot \left( {\begin{array}{*{20}{c}}
{K_\parallel ^1}\\
{K_\parallel ^2}
\end{array}} \right) = 0,
\end{array}
\label{equ89}
\end{equation}
where we have omitted all function arguments $\left( {{\bf{k}},\omega } \right)$.
This system of equations has a trivial solution, in which both current densities are zero. For a non-trivial solution the determinant of the matrix is equal to zero. Thus we get the condition for modes by letting the determinant of the matrix be equal to zero. The mode condition function becomes
\begin{equation}
\begin{array}{l}
f_{\bf{k}}^{{\rm{TM}}}\left( \omega  \right) = 1 - {e^{ - 2{\gamma ^{\left( 0 \right)}}\left( {k,\omega } \right)kd}}\\
\quad\quad \times \frac{{{\gamma ^{\left( 0 \right)}}\left( {k,\omega } \right){{\tilde \alpha }^1}_\parallel \left( {{\bf{k}},\omega } \right){\gamma ^{\left( 0 \right)}}\left( {k,\omega } \right){{\tilde \alpha }^2}_\parallel \left( {{\bf{k}},\omega } \right)}}{{\left[ {1 + {\gamma ^{\left( 0 \right)}}\left( {k,\omega } \right){{\tilde \alpha }^1}_\parallel \left( {{\bf{k}},\omega } \right)} \right]\left[ {1 + {\gamma ^{\left( 0 \right)}}\left( {k,\omega } \right){{\tilde \alpha }^2}_\parallel \left( {{\bf{k}},\omega } \right)} \right]}}
\end{array},
\label{equ90}
\end{equation}
where we have let the reference system be when the layers are at infinite separation.
For the TE modes we need the transverse fields. They are from (\ref{equ41}) and (\ref{equ42})
\begin{equation}
\begin{array}{*{20}{l}}
\begin{array}{l}
E_ \bot ^1\left( {{\bf{k}},\omega } \right) = \frac{{i\omega }}{{{c^2}}}\frac{{2\pi }}{{k{\gamma ^{\left( 0 \right)}}\left( {k,\omega } \right)}}\\
\quad \quad \quad \quad  \times \left[ {K_ \bot ^1\left( {{\bf{k}},\omega } \right) + {e^{ - {\gamma ^{\left( 0 \right)}}\left( {k,\omega } \right)kd}}K_ \bot ^2\left( {{\bf{k}},\omega } \right)} \right];
\end{array}\\
\begin{array}{l}
E_ \bot ^2\left( {{\bf{k}},\omega } \right) = \frac{{i\omega }}{{{c^2}}}\frac{{2\pi }}{{k{\gamma ^{\left( 0 \right)}}\left( {k,\omega } \right)}}\\
\quad \quad \quad \quad  \times \left[ {{e^{ - {\gamma ^{\left( 0 \right)}}\left( {k,\omega } \right)kd}}K_ \bot ^1\left( {{\bf{k}},\omega } \right) + K_ \bot ^2\left( {{\bf{k}},\omega } \right)} \right],
\end{array}
\end{array}
\label{equ91}
\end{equation}
and the induced current densities
\begin{equation}
\begin{array}{*{20}{l}}
\begin{array}{l}
K_ \bot ^1\left( {{\bf{k}},\omega } \right) = \frac{{{{\left( {\omega /ck} \right)}^2}\tilde \alpha _ \bot ^1}}{{{\gamma ^{\left( 0 \right)}}\left( {k,\omega } \right)}}\\
\quad \quad \quad \quad  \times \left[ {K_ \bot ^1\left( {{\bf{k}},\omega } \right) + {e^{ - {\gamma ^{\left( 0 \right)}}\left( {k,\omega } \right)kd}}K_ \bot ^2\left( {{\bf{k}},\omega } \right)} \right];
\end{array}\\
\begin{array}{l}
K_ \bot ^2\left( {{\bf{k}},\omega } \right) = \frac{{{{\left( {\omega /ck} \right)}^2}\tilde \alpha _ \bot ^2}}{{{\gamma ^{\left( 0 \right)}}\left( {k,\omega } \right)}}\\
\quad \quad \quad \quad  \times \left[ {{e^{ - {\gamma ^{\left( 0 \right)}}\left( {k,\omega } \right)kd}}K_ \bot ^1\left( {{\bf{k}},\omega } \right) + K_ \bot ^2\left( {{\bf{k}},\omega } \right)} \right],
\end{array}
\end{array}
\label{equ92}
\end{equation}
or on matrix form
\begin{equation}
\begin{array}{*{20}{l}}
{\left( {\begin{array}{*{20}{c}}
{\frac{{{{\left( {\omega /ck} \right)}^2}\tilde \alpha _ \bot ^1}}{{{\gamma ^{\left( 0 \right)}}\left( {k,\omega } \right)}} - 1}&{\frac{{{e^{ - {\gamma ^{\left( 0 \right)}}\left( {k,\omega } \right)kd}}{{\left( {\omega /ck} \right)}^2}\tilde \alpha _ \bot ^1}}{{{\gamma ^{\left( 0 \right)}}\left( {k,\omega } \right)}}}\\
{\frac{{{{\left( {\omega /ck} \right)}^2}\tilde \alpha _ \bot ^2}}{{{\gamma ^{\left( 0 \right)}}\left( {k,\omega } \right)}}{e^{ - {\gamma ^{\left( 0 \right)}}\left( {k,\omega } \right)kd}}}&{\frac{{{{\left( {\omega /ck} \right)}^2}\tilde \alpha _ \bot ^2}}{{{\gamma ^{\left( 0 \right)}}\left( {k,\omega } \right)}} - 1}
\end{array}} \right)}\\
{ \cdot \left( {\begin{array}{*{20}{c}}
{K_ \bot ^1\left( {{\bf{k}},\omega } \right)}\\
{K_ \bot ^2\left( {{\bf{k}},\omega } \right)}
\end{array}} \right) = 0.}
\end{array}
\label{equ93}
\end{equation}
The mode condition function for TE modes is
\begin{equation}
\begin{array}{*{20}{l}}
{f_{\bf{k}}^{{\rm{TE}}}\left( \omega  \right) = 1 - {e^{ - 2{\gamma ^{\left( 0 \right)}}\left( {k,\omega } \right)kd}}}\\
{\quad \quad  \times \frac{{\left[ {{{\left( {\omega /ck} \right)}^2}\tilde \alpha _ \bot ^1} \right]\left[ {{{\left( {\omega /ck} \right)}^2}\tilde \alpha _ \bot ^2} \right]}}{{\left[ {{\gamma ^{\left( 0 \right)}}\left( {k,\omega } \right) - {{\left( {\omega /ck} \right)}^2}\tilde \alpha _ \bot ^1} \right]\left[ {{\gamma ^{\left( 0 \right)}}\left( {k,\omega } \right) - {{\left( {\omega /ck} \right)}^2}\tilde \alpha _ \bot ^2} \right]}},}
\end{array}
\label{equ94}
\end{equation}
where we have chosen as reference system the geometry when the layers are infinitely apart.

\subsubsection{Maxwells equations and boundary conditions in fully retarded treatment\label{MEret22D}}
The normal modes can be found from self-sustained charge and current densities, from self-sustained potentials or from self-sustained fields. Here, we will find self-sustained fields. To do that one solves the MEs  in all regions of the geometry and makes use of the standard boundary conditions at all  interfaces between the regions. For the present task we need a geometry consisting of three regions and two interfaces, $1|2|3$. This geometry gives the mode condition function
\begin{equation}
{f_k} = 1 - {e^{ - 2{\gamma _2}kd}}{r_{21}}{r_{23}},
\label{equ95}
\end{equation}
where ${r_{ij}}$ is the amplitude reflection coefficient for a wave impinging on the interface between medium $i$ and $j$ from the $i$ side, $d$ is the thickness of region $2$, and ${\gamma _i} = \sqrt {1 - {{\tilde \varepsilon }_i}\left( \omega  \right){{\left( {\omega /ck} \right)}^2}} $. The function ${{{\tilde \varepsilon }_i}\left( \omega  \right)}$ is the dielectric function of medium $i$. At an interface where there is no 2D sheet the TM and TE amplitude reflection coefficients are\cite{Ser1}
\begin{equation}
r_{ij}^{TM} = \frac{{{{\tilde \varepsilon }_j}{\gamma _i} - {{\tilde \varepsilon }_i}{\gamma _j}}}{{{{\tilde \varepsilon }_j}{\gamma _i} + {{\tilde \varepsilon }_i}{\gamma _j}}},
\label{equ96}
\end{equation}
and
\begin{equation}
r_{ij}^{TE} = \frac{{\left( {{\gamma _i} - {\gamma _j}} \right)}}{{\left( {{\gamma _i} + {\gamma _j}} \right)}},
\label{equ97}
\end{equation}
respectively.  Note that ${r_{ji}} =  - {r_{ij}}$ holds for both mode types.

The amplitude reflection coefficient gets modified if there is a 2D layer at the interface. We treat the 2D layer at the interface as external to our system. The fields will then induce external surface charge and current densities at the interface. Two of the boundary conditions are enough to get the modified Fresnel coefficients. The other two are redundant. We choose the following two,
\begin{equation}
\begin{array}{l}
\left( {{{\bf{E}}_2} - {{\bf{E}}_1}} \right) \times {\bf{n}} = 0\\
\left( {{{{\bf{\tilde H}}}_2} - {{{\bf{\tilde H}}}_1}} \right) \times {\bf{n}} =  - \frac{{4\pi }}{c}{{\bf{K}}_{ext}} =  - \frac{{4\pi }}{c}\tilde {\sigma}  {\bf{n}} \times \left( {{\bf{E}} \times {\bf{n}}} \right),
\end{array}
\label{equ98}
\end{equation}
where $\tilde {\sigma} $ is the conductivity of the 2D sheet. The modified amplitude reflection coefficient for a TM mode is\cite{Ser4}
\begin{equation}
r_{ij}^{TM} = \frac{{{{\tilde \varepsilon }_j}{\gamma _i} - {{\tilde \varepsilon }_i}{\gamma _j} + 2{\gamma _i}{\gamma _j}{{\tilde \alpha }_\parallel }}}{{{{\tilde \varepsilon }_j}{\gamma _i} + {{\tilde \varepsilon }_i}{\gamma _j} + 2{\gamma _i}{\gamma _j}{{\tilde \alpha }_\parallel }}},
\label{equ99}
\end{equation}
where the polarizability of the 2D sheet is obtained from the dynamical conductivity according to (\ref{equ49}). For TM modes the tangential component of the electric field, which will induce the external current, is parallel to {\bf k}, so the longitudinal 2D dielectric function of the layer enters. The bound charges in the 2D sheet also contribute to the dynamical conductivity and the polarizability.

The modified amplitude reflection coefficient for a TE mode is\cite{Ser4}
\begin{equation}
r_{ij}^{TE} = \frac{{{\gamma _i} - {\gamma _j} + 2{{\left( {\omega /ck} \right)}^2}{{\tilde \alpha }_ \bot }}}{{{\gamma _i} + {\gamma _j} - 2{{\left( {\omega /ck} \right)}^2}{{\tilde \alpha }_ \bot }}},
\label{equ100}
\end{equation}
where the polarizability of the 2D sheet is obtained from the dynamical conductivity according to (\ref{equ49}). 
For a TE wave the electric field is perpendicular to {\bf k}, so the transverse 2D dielectric function of the layer enters. 
The bound charges in the 2D sheet also contribute to the dynamical conductivity and the polarizability.

Now, when we have geometrical structures with 2D layers at some of the interfaces we just substitute the new reflection coefficients at the proper interfaces. Note that ${r_{21}} =  - {r_{12}}$ no longer applies so one has to be careful when making the substitutions. To arrive at the starting expression one might have used this relation. For the three regions system $1|2|3$ we find the most general mode condition functions as
\begin{equation}
\begin{array}{*{20}{l}}
{f = 1 - {e^{ - 2{\gamma _2}kd}}{r_{21}}{r_{23}};}\\
{f_{\bf{k}}^{{\rm{TM}}}\left( \omega  \right)  }\\
{ = 1 - {e^{ - 2{\gamma _2}kd}}\left[ {\frac{{{{\tilde \varepsilon }_1}{\gamma _2} - {{\tilde \varepsilon }_2}{\gamma _1} + 2{\gamma _1}{\gamma _2}\tilde \alpha _L^\parallel }}{{{{\tilde \varepsilon }_1}{\gamma _2} + {{\tilde \varepsilon }_2}{\gamma _1} + 2{\gamma _1}{\gamma _2}\tilde \alpha _L^\parallel }}} \right]\left[ {\frac{{{{\tilde \varepsilon }_3}{\gamma _2} - {{\tilde \varepsilon }_2}{\gamma _3} + 2{\gamma _2}{\gamma _3}\tilde \alpha _R^\parallel }}{{{{\tilde \varepsilon }_3}{\gamma _2} + {{\tilde \varepsilon }_2}{\gamma _3} + 2{\gamma _2}{\gamma _3}\tilde \alpha _R^\parallel }}} \right];}\\
{f_{\bf{k}}^{{\rm{TE}}}\left( \omega  \right)  }\\
{ = 1 - {e^{ - 2{\gamma _2}kd}}\left[ {\frac{{{\gamma _2} - {\gamma _1} + 2{{\left( {\omega /ck} \right)}^2}\tilde \alpha _L^ \bot }}{{{\gamma _2} + {\gamma _1} - 2{{\left( {\omega /ck} \right)}^2}\tilde \alpha _L^ \bot }}} \right]\left[ {\frac{{{\gamma _2} - {\gamma _3} + 2{{\left( {\omega /ck} \right)}^2}\tilde \alpha _R^ \bot }}{{{\gamma _2} + {\gamma _3} - 2{{\left( {\omega /ck} \right)}^2}\tilde \alpha _R^ \bot }}} \right],}
\end{array}
\label{equ101}
\end{equation}
where we assumed there are 2D sheets at both interfaces. These sheets may be different so we have put the subscripts $L$ and $R$ on the polarizability of the left and right sheet, respectively.

We obtain two freestanding 2D sheets from our geometry by letting all three media be vacuum. Then the mode condition functions in (\ref{equ101}) reduce to
\begin{equation}
\begin{array}{*{20}{l}}
{f_{\bf{k}}^{{\rm{TM}}}\left( \omega  \right) = 1 - {e^{ - 2{\gamma ^{\left( 0 \right)}}kd}}\left[ {\frac{{{\gamma ^{\left( 0 \right)}}\tilde \alpha _1^\parallel \left( {k,\omega } \right)}}{{1 + {\gamma ^{\left( 0 \right)}}\tilde \alpha _1^\parallel \left( {k,\omega } \right)}}} \right]\left[ {\frac{{{\gamma ^{\left( 0 \right)}}\tilde \alpha _2^\parallel \left( {k,\omega } \right)}}{{1 + {\gamma ^{\left( 0 \right)}}\tilde \alpha _2^\parallel \left( {k,\omega } \right)}}} \right];}\\
{\begin{array}{*{20}{l}}
{f_{\bf{k}}^{{\rm{TE}}}\left( \omega  \right) = 1 - {e^{ - 2{\gamma ^{\left( 0 \right)}}kd}}\left[ {\frac{{{{\left( {\omega /ck} \right)}^2}\tilde \alpha _1^ \bot \left( {k,\omega } \right)}}{{{\gamma ^{\left( 0 \right)}} - {{\left( {\omega /ck} \right)}^2}\tilde \alpha _1^ \bot \left( {k,\omega } \right)}}} \right]}\\
{\quad \quad \quad \quad \quad \quad \quad  \times \left[ {\frac{{{{\left( {\omega /ck} \right)}^2}\tilde \alpha _2^ \bot \left( {k,\omega } \right)}}{{{\gamma ^{\left( 0 \right)}} - {{\left( {\omega /ck} \right)}^2}\tilde \alpha _2^ \bot \left( {k,\omega } \right)}}} \right]}.
\end{array}}
\end{array}
\label{equ102}
\end{equation}

\subsection{One 2D film above a substrate in fully retarded treatment\label{2DSret}}

\subsubsection{Coupled induced sources in fully retarded treatment}

When a 2D layer is placed next to a substrate there will be induced current densities at the interface. The effect of these densities can be obtained in a much simpler way using mirror images. How these mirror images look like was derived in the section \ref{Images} for a more complex system where the 2D layer is embedded in a medium. For our system the mirror-image current-densities are
\begin{equation}
\begin{array}{l}
{K_\parallel }^\prime \left( {{\bf{k}},\omega } \right) = \frac{{{\gamma _s} - {{\tilde \varepsilon }_s}\left( \omega  \right){\gamma ^{\left( 0 \right)}}}}{{{\gamma _s} + {{\tilde \varepsilon }_s}\left( \omega  \right){\gamma ^{\left( 0 \right)}}}}{K_\parallel }\left( {{\bf{k}},\omega } \right);\\
{K_ \bot }^\prime \left( {{\bf{k}},\omega } \right) = \frac{{{\gamma ^{\left( 0 \right)}} - {\gamma _s}}}{{{\gamma ^{\left( 0 \right)}} + {\gamma _s}}}{K_ \bot }\left( {{\bf{k}},\omega } \right),
\end{array}
\label{equ103}
\end{equation}
when ${K_\parallel }\left( {{\bf{k}},\omega } \right)$ and ${K_ \bot }\left( {{\bf{k}},\omega } \right)$ are the actual longitudinal and transverse, respectively, current densities in the 2D layer. These mirror images are at a distance $d$ from the interface and inside the substrate. We have suppressed the argument $\left( {k,\omega } \right)$ in all $\gamma$-functions. The coupled longitudinal current densities give rise to TM modes and the transverse to TE modes.

We start with the TM modes. The resulting longitudinal electric fields from a longitudinal current density was derived in (\ref{equ41}) and (\ref{equ42}). The field in the 2D layer is 
\begin{equation}
\begin{array}{l}
{E_\parallel }\left( {{\bf{k}},\omega } \right) =  - i2\pi \left( {k/\omega } \right){\gamma ^{\left( 0 \right)}}\left( {k,\omega } \right)\\
\quad \quad \quad \quad  \times \left[ {{K_\parallel }\left( {{\bf{k}},\omega } \right) + {e^{ - 2{\gamma ^{\left( 0 \right)}}kd}}{K_\parallel }'\left( {{\bf{k}},\omega } \right)} \right].
\end{array}
\label{equ104}
\end{equation}
We close the loop by noting that the current density in the 2D layer is the conductivity times the electric field. Thus,
\begin{equation}
\begin{array}{l}
{K_\parallel }\left( {{\bf{k}},\omega } \right) = {\tilde {\sigma}  _\parallel }\left( {{\bf{k}},\omega } \right){E_\parallel }\left( {{\bf{k}},\omega } \right)\\
 =  - \underbrace {2\pi i{\tilde {\sigma}  _\parallel }\left( {{\bf{k}},\omega } \right)\left( {k/\omega } \right)}_{{{\tilde \alpha }_\parallel }\left( {{\bf{k}},\omega } \right)}{\gamma ^{\left( 0 \right)}}\\
 \times \left[ {{K_\parallel }\left( {{\bf{k}},\omega } \right) + {e^{ - 2{\gamma ^{\left( 0 \right)}}kd}}{K_\parallel }'\left( {{\bf{k}},\omega } \right)} \right]\\
 =  - {{\tilde \alpha }_\parallel }\left( {{\bf{k}},\omega } \right){\gamma ^{\left( 0 \right)}}\left[ {1 + {e^{ - 2{\gamma ^{\left( 0 \right)}}kd}}\frac{{{\gamma _s} - {{\tilde \varepsilon }_s}\left( \omega  \right){\gamma ^{\left( 0 \right)}}}}{{{\gamma _s} + {{\tilde \varepsilon }_s}\left( \omega  \right){\gamma ^{\left( 0 \right)}}}}} \right]{K_\parallel }\left( {{\bf{k}},\omega } \right),
\end{array}
\label{equ105}
\end{equation}
where we have used (\ref{equ103}) in the last step to eliminate the mirror density. We may now use the first and last parts of this equation to find
\begin{equation}
\begin{array}{l}
\left\{ {1 + {{\tilde \alpha }_\parallel }\left( {{\bf{k}},\omega } \right){\gamma ^{\left( 0 \right)}}\left[ {1 + {e^{ - 2{\gamma ^{\left( 0 \right)}}kd}}\frac{{{\gamma _s} - {{\tilde \varepsilon }_s}\left( \omega  \right){\gamma ^{\left( 0 \right)}}}}{{{\gamma _s} + {{\tilde \varepsilon }_s}\left( \omega  \right){\gamma ^{\left( 0 \right)}}}}} \right]} \right\}\\
\quad \quad \quad \quad \quad \quad \quad \quad \quad \quad \quad \quad \quad \quad \quad \quad  \times {K_\parallel }\left( {{\bf{k}},\omega } \right) = 0.
\end{array}
\label{equ106}
\end{equation}
This has the trivial solution that the current density is zero. There are non-trivial solutions, the normal modes, if what is inside the curly brackets is zero. This leads to the mode condition function for TM modes,
\begin{equation}
\begin{array}{l}
f_{\bf{k}}^{{\rm{TM}}}\left( \omega  \right) = 1 - {e^{ - 2{\gamma ^{\left( 0 \right)}}kd}}\left[ {\frac{{{\gamma ^{\left( 0 \right)}}{{\tilde \alpha }_\parallel }\left( {{\bf{k}},\omega } \right)}}{{1 + {\gamma ^{\left( 0 \right)}}{{\tilde \alpha }_\parallel }\left( {{\bf{k}},\omega } \right)}}} \right]\\
\quad \quad \quad \quad \quad \quad \quad \quad \quad \quad \quad  \times \left[ {\frac{{{{\tilde \varepsilon }_s}\left( \omega  \right){\gamma ^{\left( 0 \right)}} - {\gamma _s}}}{{{{\tilde \varepsilon }_s}\left( \omega  \right){\gamma ^{\left( 0 \right)}} + {\gamma _s}}}} \right],
\end{array}
\label{equ107}
\end{equation}
where we have chosen as reference system the system when the 2D layer is at infinite distance from the substrate.

Now, we proceed with the TE modes. The resulting transverse electric fields from a transverse current density was derived in (\ref{equ41}) and (\ref{equ42}). The field in the 2D layer is 
\begin{equation}
{E_ \bot }\left( {{\bf{k}},\omega } \right) = \frac{{i\omega }}{{{c^2}}}\frac{{2\pi }}{{k{\gamma ^{\left( 0 \right)}}}}\left[ {{K_ \bot }\left( {{\bf{k}},\omega } \right) + {e^{ - 2{\gamma ^{\left( 0 \right)}}kd}}{K_ \bot }'\left( {{\bf{k}},\omega } \right)} \right].
\label{equ108}
\end{equation}
We close the loop by noting that the current density in the 2D layer is the conductivity times the electric field. Thus,
\begin{equation}
\begin{array}{l}
{K_ \bot }\left( {{\bf{k}},\omega } \right) = {\tilde {\sigma}  _ \bot }\left( {{\bf{k}},\omega } \right){E_ \bot }\left( {{\bf{k}},\omega } \right)\\
 = \frac{{{\omega ^2}}}{{{\gamma ^{\left( 0 \right)}}{{\left( {ck} \right)}^2}}}\underbrace {2\pi i{\tilde {\sigma}  _ \bot }\left( {{\bf{k}},\omega } \right)\left( {k/\omega } \right)}_{{{\tilde \alpha }_ \bot }\left( {{\bf{k}},\omega } \right)}\\
 \times \left[ {{K_ \bot }\left( {{\bf{k}},\omega } \right) + {e^{ - 2{\gamma ^{\left( 0 \right)}}kd}}{K_ \bot }'\left( {{\bf{k}},\omega } \right)} \right]\\
 = {{\tilde \alpha }_ \bot }\left( {{\bf{k}},\omega } \right)\frac{{{{\left( {\omega /ck} \right)}^2}}}{{{\gamma ^{\left( 0 \right)}}}}\left[ {1 + {e^{ - 2{\gamma ^{\left( 0 \right)}}kd}}\frac{{{\gamma ^{\left( 0 \right)}} - {\gamma _s}}}{{{\gamma ^{\left( 0 \right)}} + {\gamma _s}}}} \right]{K_ \bot }\left( {{\bf{k}},\omega } \right),
\end{array}
\label{equ109}
\end{equation}
where we have used (\ref{equ103}) in the last step to eliminate the mirror density. We may now use the first and last parts of this equation to find
\begin{equation}
\begin{array}{l}
\left\{ {1 - {{\tilde \alpha }_ \bot }\left( {{\bf{k}},\omega } \right)\frac{{{{\left( {\omega /ck} \right)}^2}}}{{{\gamma ^{\left( 0 \right)}}}}\left[ {1 + {e^{ - 2{\gamma ^{\left( 0 \right)}}kd}}\frac{{{\gamma ^{\left( 0 \right)}} - {\gamma _s}}}{{{\gamma ^{\left( 0 \right)}} + {\gamma _s}}}} \right]} \right\}\\
\quad \quad \quad \quad \quad \quad \quad \quad \quad \quad \quad \quad \quad \quad \quad \quad  \times {K_ \bot }\left( {{\bf{k}},\omega } \right) = 0.
\end{array}
\label{equ110}
\end{equation}
This has the trivial solution that the current density is zero. There are non-trivial solutions, the normal modes, if what is inside the curly brackets is zero. This leads to the mode condition function for TE modes,
\begin{equation}
f_{\bf{k}}^{{\rm{TE}}}\left( \omega  \right) = 1 - {e^{ - 2{\gamma ^{\left( 0 \right)}}kd}}\frac{{{{\left( {\omega /ck} \right)}^2}{{\tilde \alpha }_ \bot }\left( {{\bf{k}},\omega } \right)}}{{{\gamma ^{\left( 0 \right)}} - {{\left( {\omega /ck} \right)}^2}{{\tilde \alpha }_ \bot }\left( {{\bf{k}},\omega } \right)}}\frac{{{\gamma ^{\left( 0 \right)}} - {\gamma _s}}}{{{\gamma ^{\left( 0 \right)}} + {\gamma _s}}},
\label{equ111}
\end{equation}
where we have chosen as reference system the system when the 2D layer is at infinite distance from the substrate.

\subsubsection{Maxwells equations and boundary conditions in fully retarded treatment}
We obtain a freestanding 2D sheet above a substrate from our geometry in section \ref{MEret22D} by letting media $1$ and $2$ be vacuum and medium $3$ be the substrate. Then the mode condition functions in (\ref{equ101}) reduce to
\begin{equation}
\begin{array}{*{20}{l}}
{f_{\bf{k}}^{{\rm{TM}}}\left( \omega  \right) = 1 - {e^{ - 2{\gamma ^{\left( 0 \right)}}kd}}\left[ {\frac{{{\gamma ^{\left( 0 \right)}}{{\tilde \alpha }_\parallel }\left( {{\bf{k}},\omega } \right)}}{{1 + {\gamma ^{\left( 0 \right)}}{{\tilde \alpha }_\parallel }\left( {{\bf{k}},\omega } \right)}}} \right]\left[ {\frac{{{{\tilde \varepsilon }_s}\left( \omega  \right){\gamma ^{\left( 0 \right)}} - {\gamma _s}}}{{{{\tilde \varepsilon }_s}\left( \omega  \right){\gamma ^{\left( 0 \right)}} + {\gamma _s}}}} \right];}\\
{f_{\bf{k}}^{{\rm{TE}}}\left( \omega  \right) = 1 - {e^{ - 2{\gamma ^{\left( 0 \right)}}kd}}\left[ {\frac{{{{\left( {\omega /ck} \right)}^2}{{\tilde \alpha }_ \bot }\left( {{\bf{k}},\omega } \right)}}{{{\gamma ^{\left( 0 \right)}} - {{\left( {\omega /ck} \right)}^2}{{\tilde \alpha }_ \bot }\left( {{\bf{k}},\omega } \right)}}} \right]\left[ {\frac{{{\gamma ^{\left( 0 \right)}} - {\gamma _s}}}{{{\gamma ^{\left( 0 \right)}} + {\gamma _s}}}} \right].}
\end{array}
\label{equ112}
\end{equation}
If retardation effects are neglected there are no TE modes and the mode condition function for TM modes is reduced further to
\begin{equation}
f_{\bf{k}}^{{\rm{TM}}}\left( \omega  \right) = 1 - {e^{ - 2kd}}\left[ {\frac{{{{\tilde \alpha }_\parallel }\left( {{\bf{k}},\omega } \right)}}{{1 + {{\tilde \alpha }_\parallel }\left( {{\bf{k}},\omega } \right)}}} \right]\left[ {\frac{{{{\tilde \varepsilon }_s}\left( \omega  \right) - 1}}{{{{\tilde \varepsilon }_s}\left( \omega  \right) + 1}}} \right],
\label{equ113}
\end{equation}
which is fully in line with what we obtained starting from the non-retarded theory.

\section{Summary and conclusions\label{Sum}}

We have derived the expressions for the dispersion interaction in systems containing 2D sheets in several different complementary ways, all with a condensed matter physics perspective. The geometries considered were two free standing 2D sheets and one 2D sheet above a substrate. The derivations were either performed using many-body theory with diagrammatic perturbation expansions based on Feynman diagrams or were based on electromagnetic normal modes. The normal modes were generated in several different ways. The zero-point energy is a very important concept in the normal mode derivation of the dispersion interaction. It is very interesting to note that the same results are obtained using many-body theory without invoking zero-point energy.

\end{document}